\documentclass[aps,amsmath,amssymb,showpacs,superscriptaddress,twocolumn,prb]{revtex4}
\usepackage{graphicx}
\usepackage{epsfig}
\usepackage{ulem}
\usepackage{amsmath}
\usepackage{amssymb}
\usepackage{physics}
\usepackage{mathtools}

\begin{document}

\title{Electronic structure and carrier transfer in \\ B-DNA monomer polymers and dimer polymers: \\ Stationary and time-dependent aspects of wire model vs. extended ladder model}
%\thanks{Stationary and time-dependent aspects}

\date{\today}% It is always \today, today,
             %  but any date may be explicitly specified

\author{K. Lambropoulos}
\affiliation{National and Kapodistrian University of Athens, Faculty of Physics, Panepistimiopolis, 15784 Zografos, Athens, Greece}

\author{M. Chatzieleftheriou}
\thanks{Current affiliation: University of Copenhagen, Niels Bohr Institute, Blegdamsvej 17, DK-2100 Copenhagen, Denmark.}

\author{A. Morphis}
\affiliation{National and Kapodistrian University of Athens, Faculty of Physics, Panepistimiopolis, 15784 Zografos, Athens, Greece}

\author{K. Kaklamanis}
\affiliation{National and Kapodistrian University of Athens, Faculty of Physics,
Panepistimiopolis, 15784 Zografos, Athens, Greece}

\author{R.Lopp}
\thanks{Current affiliation: Georg-August-Universit\"{a}t G\"{o}ttingen, Fakult\"{a}t f\"{u}r Physik,
Friedrich-Hund-Platz 1 D-37077 G\"{o}ttingen, Germany.
%{\color{red} Perimeter Institute for Theoretical Physics, Waterloo, Ontario, Canada}
}

\author{M. Theodorakou}
\affiliation{National and Kapodistrian University of Athens, Faculty of Physics,
Panepistimiopolis, 15784 Zografos, Athens, Greece}

\author{M. Tassi}
\affiliation{National and Kapodistrian University of Athens, Faculty of Physics,
Panepistimiopolis, 15784 Zografos, Athens, Greece}

\author{C. Simserides}
\email{csimseri@phys.uoa.gr}
\homepage{http://users.uoa.gr/~csimseri/physics_of_nanostructures_and_biomaterials.html}
\affiliation{National and Kapodistrian University of Athens, Faculty of Physics,
Panepistimiopolis, 15784 Zografos, Athens, Greece}


% 82.       Physical chemistry and chemical physics
% 82.39.-k 	Chemical kinetics in biological systems
% 82.39.Jn 	Charge (electron, proton) transfer in biological systems
% 82.39.Pj 	Nucleic acids, DNA and RNA bases

% 87. 	    Biological and medical physics
% 87.14.-g 	Biomolecules: types
% 87.14.gk 	DNA
% 87.15.-v 	Biomolecules: structure and physical properties
% 87.15.A- 	Theory, modeling, and computer simulation

% 72. 	    Electronic transport in condensed matter
% 72.80.-r 	Conductivity of specific materials
% 72.80.Le 	Polymers; organic compounds (including organic semiconductors)

% 73. 	    Electronic structure and electrical properties of surfaces, interfaces, thin films, and low-dimensional structures
% 73.63.-b 	Electronic transport in nanoscale materials and structures

\begin{abstract}
We employ two Tight-Binding (TB) approaches to study the electronic structure and hole or electron transfer in B-DNA monomer polymers and dimer polymers made up of $N$ monomers (base pairs): (I) at the base-pair level, using the on-site energies of base pairs and the hopping integrals between successive base pairs, i.e., a wire model and (II) at the single-base level, using the on-site energies of the bases and the hopping integrals between neighboring bases, i.e., an \textit{extended} ladder model since we also include diagonal hoppings. We solve a system of $MD$ (``matrix dimension'') coupled equations [(I) $MD$ = $N$,  (II) $MD$ = $2N$] for the time-independent problem, and a system of $MD$ coupled $1^\text{st}$ order differential equations for the time-dependent problem. We study the HOMO and the LUMO eigenspectra, the occupation probabilities, the Density of States (DOS) and the HOMO-LUMO gap as well as the mean over time probabilities to find the carrier at each site [(I) base pair or (II) base)], the Fourier spectra, which reflect the frequency content of charge transfer (CT) and the pure mean transfer rates from a certain site to another. The two TB approaches give coherent, complementary aspects of electronic properties and charge transfer in B-DNA monomer polymers and dimer polymers.
\end{abstract}

\maketitle
%%%%%%%%%%%%%%%%%%%%%%%%%%%%%%%%%%%%%%%%%%%%%%%%%%
\section{\label{sec:introduction} Introduction} %%
%%%%%%%%%%%%%%%%%%%%%%%%%%%%%%%%%%%%%%%%%%%%%%%%%%
Today, remarkable parts of the physical, chemical, biological and medical communities as well as a broad spectrum of other scientists and engineers is interested in charge transfer (CT) in biological systems. CT is the basis of many biological processes e.g. in various proteins~\cite{Page:2003} including metalloproteins~\cite{GrayWinkler:2010} and enzymes~\cite{Moser:2010}, with medical and bioengineering applications \cite{Artes:2014,Kannan:2009}.
CT plays a central role in DNA damage and repair \cite{Dandliker:1997,Rajski:2000,Giese:2006}.
CT might also be an indicator to discriminate between pathogenic and non-pathogenic mutations at an early stage \cite{Shih:2011}.

DNA plays a key role in the development, function and reproduction of living organisms, because  the sequence of its bases (adenine, guanine, thymine, cytosine) carries their genetic code, hence, its study is usually associated with molecular biology and genetics. However, its remarkable properties have spurred in recent years the interest of a broad interdisciplinary community. From the perspective of physics, its electronic structure and its CT properties are studied with the aim of understanding  its biological functions as well as its potential applications in nanotechnology (e.g. nanosensors, nanocircuits, molecular wire).

At least for twenty years, we try to understand carrier movement through DNA~\cite{FinkSchoenenberger:1999,Porath:2000,Storm:2001,Yoo:2001,Xu:2004,Cohen:2005,TSK:2005}.
Today, we know that many external factors related to the environment, like aqueousness and presence of counterions, extraction process, conduct quality with electrodes, purity, substrate and so on, influence carrier movement. This leads to the need of a deeper understanding of endogenous factors affecting carrier movement in DNA, like base-pair sequence and geometry. Maybe the most important endogenous factor is the base-pair sequence, to which this article is devoted, too.

Additionally, we have to discriminate between the words \textit{transport} (usually implying the use of electrodes), \textit{transfer}, and \textit{migration} (a transfer over rather long distances). The carriers (electrons or holes) can be either inserted via electrodes or generated by UV irradiation and by chemical reduction or oxidation. Moreover, although unbiased charge transfer in DNA nearly vanishes after 10 to 20 nm~\cite{Simserides:2014, LChMKTS:2015}, DNA still remains a promising candidate as an electronic component in molecular electronics, e.g. as a short molecular wire~\cite{Wohlgamuth:2013}. Favoring geometries and base-pair sequences have still to be explored e.g. incorporation of sequences serving as molecular rectifiers, using non-natural bases or using the triplet acceptor anthraquinone for hole injection~\cite{LewisWasielewski:2013}. Structural fluctuations could be another important factor which influences quantum transport through DNA molecular wires~\cite{Gutierrez:2010}. Finally, the carrier transfer rate through DNA can be manipulated by chemical modification~\cite{KawaiMajimaBook:2015}.

On theoretical side, both \textit{ab initio} calculations \cite{YeShen:2000,YeJiang:2000, Barnett:2003,Artacho:2003,Adessi:2003,MehrezAnantram:2005,Voityuk:2008}
and model Hamiltonians~\cite{Cuniberti:2002,Roche-et-al:2003,Roche:2003,Palmero:2004,Yamada:2004,ApalkovChakraborty:2005,Klotsa:2005,Shih:2008,Joe:2010}
try to interpret the diversity of experimental results and ascertain the underlying CT mechanism.
The former can provide a more detailed description, but are currently limited to very  short segments, while the latter are much less detailed but allowing to address systems of realistic length, grasping hopefully the underlying physics \cite{Albuquerque:2014}.
Here we study rather long B-DNA segments, hence we adopt the latter approach.

Specifically, we employ two Tight-Binding (TB) approaches.
TB I is very simple, it is an approach at the base-pair (bp) level. We need the on-site energies of base pairs and the hopping integrals between successive base pairs. In other words, TB I is a \textit{wire} model \cite{Cuniberti:2007}.
TB II is an approach at the single-base (sb) level. We need the on-site energies of bases and the hopping integrals between neighboring bases. We also include diagonal hoppings, in that sense, TB II is an  \textit{extended ladder} model~\cite{Albuquerque:2014}.
The inclusion of diagonal hoppings is crucial in some cases as will become evident below.
With these two TB models we study the electronic structure and hole or electron transfer in B-DNA monomer polymers and dimer polymers. This means that we call \textit{monomer} a B-DNA base pair and study polymers made of $N$ monomers, with repetition unit one or two monomers. To this end, we shall see below, we have to solve a system of $MD$ (``matrix dimension'') coupled equations for the time-independent problem, and a system of $MD$ coupled $1^\text{st}$ order differential equations for the time-dependent problem. In TB I $MD$ = $N$, while in TB II $MD$ = $2N$.
In this article, we study HOMO and LUMO eigenspectra and the relevant Density of States (DOS) as well as the mean over time probabilities to find the carrier at each site, which is a base pair for TB I and a base for TB II. We are also interested in the frequency content of carrier movements, hence, we analyze the Fourier spectra, too. The pure mean transfer rate from a certain site to another describes the easiness of CT; it gives us a measure of how much of the carrier is transferred and  also of how fast this process is. Our two TB approaches give coherent, complementary aspects of electronic properties and charge transfer in these B-DNA monomer polymers and dimer polymers.

The rest of the paper is organized as follows: In Section ~\ref{sec:theory} we delineate the basic theory behind the time-independent (Subsection~\ref{subsec:TIproblem}) and the time-dependent (Subsection~\ref{subsec:TDproblem}) problem.
In Section~\ref{sec:Results} we discuss our results for
eigenspectra and eigenvectors (Subsection~\ref{subsec:eigenSpectraVectors}),
the density of states (Subsection~\ref{subsec:DOS}),
the HOMO-LUMO gaps (Subsection~\ref{subsec:HLGaps}),
the mean over time probabilities to find the carrier at each site (Subsection~\ref{subsec:MeanProbs}),
the CT frequency content (Subsection~\ref{subsec:FrequencyContent}), and
the pure mean transfer rates (Subsection~\ref{subsec:meantransferrates}).
In Section~\ref{sec:Conclusion} we state our conclusions.

%%%%%%%%%%%%%%%%%%%%%%%%%%%%%%%%%%%%%%%
\section{\label{sec:theory} Theory} %%%
%%%%%%%%%%%%%%%%%%%%%%%%%%%%%%%%%%%%%%%
Let us begin with some notations. We call \textit{monomer} a B-DNA base pair.
We denote a system of two successive monomers by YX, according to the convention
\begin{equation}
\begin{tabular}{c|ccc} \label{DNAnotation}
	         & $\sigma$=1 & & $\sigma$=2  \\ \hline
	 \vdots  & $5'$       & & $3'$        \\
	 $\mu$   & Y          &-& Y$_\text{compl}$ \\
	 $\mu+1$ & X          &-& X$_\text{compl}$ \\
	 \vdots  & $3'$       & & $5'$
\end{tabular}
\end{equation}
for the B-DNA strands orientation. X$_{\text{compl}}$ (Y$_{\text{compl}}$) is the complementary base of X (Y). The base pair X-X$_{\text{compl}}$ is separated and twisted by 3.4 {\AA} and $36^{\circ}$, respectively, relatively to the base pair Y-Y$_{\text{compl}}$, around the B-DNA growth axis. We call $\mu$ the monomer index, with $\mu = 1,2,\dots,N$, and $\sigma$ the strand index ($\sigma=1$ for the strand with $5'$-$3'$ directionality, $\sigma=2$ for the strand with $3'$-$5'$ directionality). Further, we define the base index $\beta(\mu,\sigma)$, $\beta = 1,2,\dots,2N$, according to the expression $\beta = 2(\mu-1) + \sigma$. Schematically,
\begin{equation*}
\begin{tabular}{c|c||c} \label{indices}
	$\mu$ & $\sigma$ & $\beta$  \\ \hline
	1     & 1      &  1  \\
	1     & 2      &  2  \\
	2     & 1      &  3  \\
	2     & 2      &  4  \\
	\vdots & \vdots & \vdots
\end{tabular}
\end{equation*}
In this work, we study all possible periodic B-DNA segments of the form YXYX$\dots$, consisting of $N$ monomers, i.e., monomer polymers and dimer polymers.
There are three types of such polymers: \\
(type $\alpha'$) poly(dG)-poly(dC), poly(dA)-poly(dT), \\
(type $\beta'$)  GCGC\dots, CGCG\dots, ATAT\dots, TATA\dots, and \\
(type $\gamma'$) ACAC\dots, CACA\dots, TCTC\dots, CTCT\dots, AGAG\dots, GAGA\dots, TGTG\dots,  GTGT\dots. \\
We employ two Tight-Binding (TB) approaches to study the electronic structure and single carrier transfer in such B-DNA polymers, under the hypothesis that an extra hole or electron travels through HOMOs or LUMOs, respectively.
(I) Within TB I (description at the base-pair level),
we use the HOMO/LUMO on-site energies of base pairs and the HOMO/LUMO hopping integrals between successive base pairs.
The TB parameters for TB I are the same as in Refs.~\cite{Simserides:2014,LChMKTS:2015,LKGS:2014,Lambropoulos:2015}.
(II) Within TB II (description at the single-base level),
we use the HOMO/LUMO on-site energies of bases and the HOMO/LUMO hopping integrals between (a) two successive bases on the same strand, (b) complementary bases that constitute a monomer, and (c) diagonally located bases of successive monomers, in the directions $5'$-$5'$ and $3'-3'$. The TB parameters for TB II are taken from Ref.~\cite{HKS:2010-2011}. In other words, within TB I, a monomer is considered as a single site, characterized by the index $\mu$, while, within TB II, a base is considered as a single site, characterized by the index $\beta$.
Below, we use a generic site index $j$, $j=1,2,\dots,MD$, where,
$j=\mu$ and $MD = N$, for TB I, while,
$j=\beta$ and $MD = 2N$, for TB II.
$MD$ denotes ``matrix dimension''.

%%%%%%%%%%%%%%%%%%%%%%%%%%%%%%%%%%%%%%%%%%%%%%%%%%%%%%%%%%%%%%%%%%%%%%%%%%%%%%%%%%%%%%
\subsection{\label{subsec:TIproblem} Stationary States - Time-independent problem} %%%
%%%%%%%%%%%%%%%%%%%%%%%%%%%%%%%%%%%%%%%%%%%%%%%%%%%%%%%%%%%%%%%%%%%%%%%%%%%%%%%%%%%%%%
The HOMO/LUMO Hamiltonian of a given B-DNA segment can be written as
\begin{equation} \label{Hamiltonian}
\hat{H} = \sum_{j=1}^{MD} E^{s(j)} \ketbra{j}{j} + \sum_{<j,j'>} t^{s(j,j')} \big(\ketbra{j}{j'} + h.c.\big),
\end{equation}
where $E^{s(j)}$ is the HOMO/LUMO on-site energy of the $j$-th site [base pair ($bp$) for TB I or base ($b$) for TB II], and $t^{s(j,j')} (= t^{s(j',j)*})$ is the HOMO/LUMO hopping integral between the sites $j$ and $j'$. $<j,j'>$ denotes summation over all relevant neighbors. The neighboring sites which are taken into account for each TB approach are described above.
For TB I (\textit{wire} model), the Hamiltonian can be written as
\begin{equation} \label{WireHamiltonian}
\hat{H}_\text{W} =
\sum_{\mu=1}^{N} E^{bp(\mu)} \ketbra{\mu}{\mu} +
\bigg( \sum_{\mu=1}^{N-1} t^{bp(\mu,\mu+1)} \ketbra{\mu}{\mu+1} + h.c. \bigg).
\end{equation}
For TB II (\textit{extended ladder} model), the Hamiltonian can be written as
\begin{align} \label{ExtendedLadderHamiltonian}
\hat{H}_\text{EL} & = \sum_{\beta=1}^{MD} E^{b(\beta)} \ketbra{\beta}{\beta} \\ \nonumber
& + \bigg( \sum_{\beta=1}^{MD-2} t^{b(\beta,\beta+2)} \ketbra{\beta}{\beta+2} + h.c. \bigg) \\ \nonumber
& + \bigg( \sum_{\beta=1, odd}^{MD-1} t^{b(\beta,\beta+1)} \ketbra{\beta}{\beta+1} + h.c. \bigg) \\ \nonumber
& + \bigg( \sum_{\beta=1, odd}^{MD-3} t^{b(\beta,\beta+3)} \ketbra{\beta}{\beta+3} + h.c. \bigg) \\ \nonumber
& + \bigg( \sum_{\beta=2, even}^{MD-2} t^{b(\beta,\beta+1)} \ketbra{\beta}{\beta+1} + h.c. \bigg), \\ \nonumber
\end{align}
where the second term represents intra-strand,
the third intra-base-pair, the fourth inter-strand $5'$-$5'$ and the fifth inter-strand $3'$-$3'$ hoppings.
In the context of TB, we suppose that $\braket{j}{j'} = \delta_{jj'}.$

The HOMO/LUMO state of the segment can be expressed as
\begin{equation} \label{TIDNA}
\ket{\text{DNA}} = \sum_{j=1}^{MD} v_j \ket{j}.
\end{equation}
Substituting Eqs. \eqref{Hamiltonian} and \eqref{TIDNA} to the time-independent Schr\"odinger equation
\begin{equation} \label{TISchr}
\hat{H} \ket{\text{DNA}} = E \ket{\text{DNA}},
\end{equation}
we arrive to a system of $MD$ coupled equations. Within TB I, the system is of the form
\begin{equation} \label{TIsystembp}
E u_{\mu} = E^{bp(\mu)} v_{\mu} + t^{bp(\mu, \mu+1)} v_{\mu+1} + t^{bp(\mu, \mu-1)} v_{\mu-1},
\end{equation}
for $\mu$ even or odd, while, within TB II, the system is of the form
\begin{subequations} \label{TIsystemsb}
\begin{align}
E v_\beta =& t^{b(\beta,\beta-2)} v_{\beta-2} + t^{b(\beta,\beta-1)} v_{\beta-1} + E^{b(\beta)} v_\beta + \nonumber \\ &t^{b(\beta,\beta+1)} v_{\beta+1} + t^{b(\beta,\beta+2)} v_{\beta+2} + t^{b(\beta,\beta+3)} v_{\beta+3},
\end{align}
for $\beta$ odd, i.e., for the bases of strand 1, and
\begin{align}
E v_\beta =& t^{b(\beta,\beta-3)} v_{\beta-3} + t^{b(\beta,\beta-2)} v_{\beta-2} + t^{b(\beta,\beta-1)}  v_{\beta-1} + \nonumber \\ &E^{b(\beta)} v_\beta + t^{b(\beta,\beta+1)} v_{\beta+1} + t^{b(\beta,\beta+2)} v_{\beta+2},
\end{align}
for $\beta$ even, i.e., for the bases of strand 2.
\end{subequations}
Eqs. \eqref{TIsystembp} and \eqref{TIsystemsb} are equivalent  to the eigenvalue-eigenvector problem
\begin{equation} \label{eigenvproblem}
H\vec{v} = E \vec{v},
\end{equation}
where $H$ is the hamiltonian matrix of order $MD$, composed of the TB parameters $E^s$ and $t^s$, and $\vec{v}$ is the vector matrix composed of the coefficients $v_j$. For the segments studied in this work, within TB I, $H$ is either a tridiagonal symmetric Toeplitz matrix of order $N$ for type \textit{$\alpha'$} polymers, or a tridiagonal symmetric 2-Toeplitz matrix of order $N$ for type \textit{$\beta'$} and type $\gamma'$ polymers. Within TB II, $H$ is a heptadiagonal 4-Toeplitz matrix of order $2N$, or, seen another way, a tridiagonal block Toeplitz matrix of order $\frac{N}{2}$, with blocks of order 4. The diagonalization of $H$ leads to the determination of the HOMO/LUMO eigenenergy spectra (\textit{eigenspectra}), $\{E_k\}$, $k =1,2,\dots,MD$, for which we suppose that $E_1<E_2<\dots<E_{MD}$, as well as to the determination of the occupation probabilities for each eigenstate, $\abs{v_{jk}}^2$, where $v_{jk}$ is the $j$-th component of the $k$-th eigenvector. $\{v_{jk}\}$ are normalized, and their linear independence is checked in all cases.

Having determined the eigenspectra, we can compute the Density of States (DOS), generally given by
\begin{equation} \label{DOS}
g(E) = \sum_{k}^{MD} \delta(E-E_k).
\end{equation}
%{\color{red}  Let us call $z$-axis the B-DNA growth axis. Let us call $q$ the perpendicular to the $z$-axis plane which bisects the B-DNA sequence. Let us call O the intersection of $z$-axis and $q$-plane. The projections of the first base pair A$_1$-B$_1$ and the last base pair A$_N$-B$_N$ on $q$, A$_1'$-B$_1'$ and A$_N'$-B$_N'$, respectively, define the angle $\widehat{\text{A}_1'\text{OA}_N'} = \widehat{\text{B}_1'\text{OB}_N'} = \theta$. Let us define $y$-axis as lying on $q$-plane and bisecting $\theta$. Let us suppose that we look at the B-DNA segment from the top of $z$-axis. Then, the rotation $C_2^x$, i.e., $180^{\circ}$ along the $x$-axis (equivalent with viewing the B-DNA sequence from the bottom of $z$-axis), reflects the hamiltonian matrix $H$ of the segment along its main antidiagonal.}
Changing the view of a B-DNA segment from one (e.g. top) to the other (e.g. bottom) side of the growth axis, reflects the hamiltonian matrix $H$ of the segment on its main antidiagonal.
This reflected Hamiltonian, $H^{\text{equiv}}$, describes the \textit{equivalent polymer}. $H$ and $H^{\text{equiv}}$ are connected by the similarity transformation $H^{\text{equiv}} = P^{-1}HP$, where $P(=P^{-1})$ is the unit antidiagonal matrix of order $MD$.
Therefore, $H$ and $H^{\text{equiv}}$ have identical eigenspectra (hence the equivalent polymers' DOS is identical) and their eigenvectors are connected by $v_{jk} = v_{(MD-j+1)k}^{\text{equiv}}$. For the types of B-DNA polymers studied in this work,
\begin{equation} \label{Eq:equiv2}
\text{equiv(\text{YX\dots})} =
\begin{cases}
\text{Y}_{\text{compl}}\text{X}_{\text{compl}}\dots, & \text{for odd}  \ N \\
\text{X}_{\text{compl}}\text{Y}_{\text{compl}}\dots, & \text{for even} \ N
\end{cases}
\end{equation}
For example, for $N$ odd, ACAC{\dots} $\equiv$ TGTG{\dots}, while, for $N$ even, ACAC{\dots} $\equiv$ GTGT{\dots}.

%%%%%%%%%%%%%%%%%%%%%%%%%%%%%%%%%%%%%%%%%%%%%%%%%%%%%%%%%%%%%%%%
\subsection{\label{subsec:TDproblem} Time-dependent problem} %%%
%%%%%%%%%%%%%%%%%%%%%%%%%%%%%%%%%%%%%%%%%%%%%%%%%%%%%%%%%%%%%%%%
To describe the spatiotemporal evolution of an extra carrier (hole/electron), inserted or created (e.g. by oxidation/reduction) in a B-DNA segment, we consider the HOMO/LUMO state of the segment as
\begin{equation} \label{TDDNA}
\ket{\text{DNA}(t)} = \sum_{j=1}^{MD}C_j(t) \ket{j},
\end{equation}
where $\abs{C_j(t)}^2$ is the probability of finding the carrier at the $j$-th site at time $t$. Substituting Eqs. \eqref{Hamiltonian} and \eqref{TDDNA} to the time-dependent Schr\"odinger equation
\begin{equation} \label{TDSchr}
\hat{H} \ket{\text{DNA(t)}} = i\hbar \pdv{t}\ket{\text{DNA(t)}},
\end{equation}
we arrive at a system of $MD$ coupled differential equations of 1$^\text{st}$ order. With TB I, the system is of the form
\begin{equation} \label{TDsystembp}
i\hbar \dv{C_\mu}{t} = E^{bp(\mu)} C_{\mu} + t^{bp(\mu, \mu+1)} C_{\mu+1} + t^{bp(\mu, \mu-1)} C_{\mu-1},
\end{equation}
for $\mu$ even or odd. With TB II, the system is of the form
\begin{subequations} \label{TDsystemsb}
\begin{align}
i\hbar \dv{C_\beta}{t} =& t^{b(\beta,\beta-2)} C_{\beta-2} + t^{b(\beta,\beta-1)} C_{\beta-1} + E^{b(\beta)} C_\beta + \nonumber \\ &t^{b(\beta,\beta+1)} C_{\beta+1} + t^{b(\beta,\beta+2)} C_{\beta+2} + t^{b(\beta,\beta+3)} C_{\beta+3},
\end{align}
for $\beta$ odd, and
\begin{align}
i\hbar \dv{C_\beta}{t} =& t^{b(\beta,\beta-3)} C_{\beta-3} + t^{b(\beta,\beta-2)} C_{\beta-2} + t^{b(\beta,\beta-1)}  C_{\beta-1} + \nonumber \\ &E^{b(\beta)} C_\beta + t^{b(\beta,\beta+1)} C_{\beta+1} + t^{b(\beta,\beta+2)} C_{\beta+2},
\end{align}
for $\beta$ even.
\end{subequations}
Eqs. \eqref{TDsystembp} and \eqref{TDsystemsb} are equivalent  to a 1$^\text{st}$ order matrix differential equation of the form
\begin{equation} \label{Matrixdiffeq}
\dot{\vec{C}}(t) = -\frac{i}{\hbar}H \vec{C}(t),
\end{equation}
where $\vec{C}(t)$ is a vector matrix composed of the coefficients $C_j(t), \;\; j = 1, 2, \dots, MD$. Eq. \eqref{Matrixdiffeq} can be solved with the eigenvalue method, i.e., by looking for solutions of the form $\vec{C}(t) = \vec{v} e^{-\frac{i}{\hbar}Et} \Rightarrow \dot{\vec{C}}(t) = -\frac{i}{\hbar}E \vec{v}e^{-\frac{i}{\hbar}Et}$. Hence, Eq. \eqref{Matrixdiffeq} leads to the eigenvalue problem of Eq. \eqref{eigenvproblem}, that is, $H\vec{v} = E \vec{v}$. Having determined the eigenvalues and eigenvectors of $H$, the general solution of Eq. \eqref{Matrixdiffeq} is
\begin{equation} \label{generalsolution}
\vec{C}(t) = \sum_{k=1}^{MD} c_k \vec{v}_k e^{-\frac{i}{\hbar}E_kt},
\end{equation}
where the coefficients $c_k$ are determined from the initial conditions.
In particular, if we define the $MD \times MD$ eigenvector matrix $V$, with elements $v_{jk}$, then it can be shown that the vector matrix $\vec{c}$, composed of the coefficients $c_k, \;\; k = 1, 2, \dots, MD$, is given by the expression
\begin{equation} \label{c_kdef}
\vec{c} = V^T\vec{C}(0).
\end{equation}
Suppose that the extra carrier is placed at the $l$-th site, that is $C_l(0) = 1$, $C_j(0) = 0, \forall j \neq l$. Then,
\begin{equation} \label{cmatrix}
\vec{c} = \begin{bmatrix}
u_{l1}\\\vdots\\u_{lk}\\\vdots\\u_{lMD}
\end{bmatrix}.
\end{equation}
In other words, the coefficients $c_k$ are given by the row of the eigenvector matrix which corresponds to the site the carrier is initially placed at. In this work, within TB I, we choose $l =1$, i.e., we initially place the extra carrier at the first monomer (\textit{initial condition}), and, within TB II, we choose either $l=1$ (\textit{initial condition 1}) or $l=2$ (\textit{initial condition 2}), i.e., we initially place the extra carrier at each base of the first monomer.

From Eq. \eqref{generalsolution} it follows that the probability to find the extra carrier at the $j$-th site of a B-DNA segment is
\begin{equation} \label{probabilities}
\abs{C_j(t)}^2 = \sum_{k=1}^{MD} c_k^2v_{jk}^2 + 2 \sum_{k=1}^{MD} \sum_{\substack{k'=1 \\k'<k}}^{MD} c_k c_{k'} v_{jk} v_{jk'} \cos(2\pi f_{kk'}t),
\end{equation}
where
\begin{equation} \label{fandT}
f_{kk'} = \frac{1}{T_{kk'}} = \frac{E_k-E_{k'}}{h}, \; \forall k > k',
\end{equation}
are the frequencies ($f_{kk'}$) or periods ($T_{kk'}$) involved in charge transfer. If $M$ is the number of discrete eigenenergies, then, the number of different $f_{kk'}$ or $T_{kk'}$ involved in CT is $S= {M \choose 2} = \frac{M!}{2!(M-2)!} = \frac{M(M-1)}{2}$.
If there are no degenerate eigenenergies (which holds for all cases studied here,  but e.g. does not hold in TB I for \textit{cyclic} type $\alpha'$ polymers~\cite{LChMKTS:2015}), then $M = MD$. If eigenenergies are symmetric relative to some central value, then, $S$ decreases (there exist degenerate $f_{kk'}$ or $T_{kk'}$). Specifically, in that case, $S = \frac{M^2}{4}$, for even $M$ and $S = \frac{M^2-1}{4}$ for odd $M$.

From Eq. \eqref{probabilities} it follows that the mean over time probability to find the extra carrier at the $j$-th site is
\begin{equation} \label{meanprobabilities}
\expval{\abs{C_j(t)}^2} = \sum_{k=1}^{MD} c_k^2v_{jk}^2.
\end{equation}

Furthermore, from Eq. \eqref{probabilities} it can be shown that the one-sided Fourier amplitude spectrum that corresponds to the probability $\abs{C_j(t)}^2$ is given by
\begin{equation} \label{Fourierspectra}
\resizebox{\hsize}{!}{$\displaystyle
\abs{\mathcal{F}_j(f)} = \sum_{k=1}^{MD} c_k^2v_{jk}^2 \delta(f) + 2 \sum_{k=1}^{MD} \sum_{\substack{k'=1 \\ k'<k}}^{MD} \abs{c_k c_{k'} v_{jk} v_{jk'}} \delta(f-f_{kk'}).$}
\end{equation}

A quantity that can be defined to estimate the transfer rate, i.e., simultaneously, the magnitude of charge transfer and the time scale of the phenomenon, is the \textit{pure} mean transfer rate
\begin{equation} \label{pmtr}
k_{j'j} = \frac{\expval{\abs{C_j(t)}^2}}{t_{j'j}}.
\end{equation}
$t_{j'j}$ is the \textit{mean transfer time}, i.e.,
having placed the carrier initially at site $j'$,
the time it takes for the probability to find the extra carrier at site $j$,
$\abs{C_j(t)}^2$, to become equal to its mean value, $\expval{\abs{C_j(t)}^2}$, for the first time. For the pure mean transfer rates it holds
\begin{eqnarray} \label{kproperties}
& k_{j'j} = k_{jj'} = \nonumber \\
& k_{(MD-j'+1)(MD-j+1)}^{\text{equiv}} = k_{(MD-j+1)(MD-j'+1)}^{\text{equiv}}.
\end{eqnarray}

%%%%%%%%%%%%%%%%%%%%%%%%%%%%%%%%%%%%%%%%
\section{\label{sec:Results} Results} %%
%%%%%%%%%%%%%%%%%%%%%%%%%%%%%%%%%%%%%%%%

%%%%%%%%%%%%%%%%%%%%%%%%%%%%%%%%%%%%%%%%%%%%%%%%%%%%%%%%%%%%%%%%%%%%%%%%%%%%%%%%%%%%%%%%%%%%
\subsection{\label{subsec:eigenSpectraVectors} Eigenspectra and occupation probabilities} %%
%%%%%%%%%%%%%%%%%%%%%%%%%%%%%%%%%%%%%%%%%%%%%%%%%%%%%%%%%%%%%%%%%%%%%%%%%%%%%%%%%%%%%%%%%%%%
Let us start by saying that within TB II (TB I), we take the HOMO or LUMO eigenenergies of bases (base pairs) as the on-site energy of a hole or an electron on a base (base pair).
Using the HOMO or LUMO energies of the bases that constitute a base pair, we can estimate the HOMO or LUMO energy of the base pair~\cite{HKS:2010-2011}.
Specifically, supposing that
$\ket{\psi_{bp}} = \mathcal{C}_1 \ket{\psi_{b1}} + \mathcal{C}_2 \ket{\psi_{b2}}$,
and taking the time-independent Schr\"{o}dinger equation
$\hat{H}\ket{\psi_{bp}} = E \ket{\psi_{bp}}$
we find that the base pair eigenenergies are
$E_{1,2} = \frac{E_{b1}+E_{b2}}{2} \pm \sqrt{(\frac{E_{b1}-E_{b2}}{2})^2 + t^2}$,
where $E_{b1}$ and $E_{b2}$ are the on-site energies of the bases and
$t = \langle \psi_{b1} | \hat{H} | \psi_{b2} \rangle$ is the intra-base-pair hopping integral i.e. between the two bases that constitute a base pair.
However,
due to the weak hydrogen bonding between the bases that constitute a base pair,
$t$ is very small~\cite{HKS:2010-2011}, of the order of 10 meV.
As a result, practically, $E_{1,2} \approx E_{b1}, E_{b2}$
(with accuracy of 1 meV). Hence, we make the \textit{Observation}:
Approximately,
the HOMO of the base pair is the highest HOMO of the two bases and
the LUMO of the base pair is the lowest  LUMO of the two bases.
This is expressed in Table~\ref{Table:on-site_Energies}, where
we show all energies in eV with accuracy of 0.1 eV.
\begin{table}[h!]
\caption{On-site HOMO / LUMO energies of B-DNA bases and base pairs~\cite{HKS:2010-2011}. All energies are given in eV.}
\begin{tabular}{|r|rr|rr|} \hline
base      & Adenine & Thymine & Guanine   & Cytosine \\ \hline
$E_H^b$   & $-8.3$  & $-9.0$  &   $-8.0$  &   $-8.8$ \\
$E_L^b$   & $-4.4$  & $-4.9$  &   $-4.5$  &   $-4.3$ \\
$E_g$     &   3.9   &   4.1   &      3.5  &     4.5  \\ \hline
base pair &  \multicolumn{2}{c|}{A-T} & \multicolumn{2}{c|}{G-C} \\ \hline
$E_H^{bp}$&  $-8.3$ &         &  $-8.0$   &          \\
$E_L^{bp}$&         & $-4.9$  &  $-4.5$   &          \\
$E_g$     &  \multicolumn{2}{c|}{3.4} & \multicolumn{2}{c|}{3.5} \\ \hline
\end{tabular}
\label{Table:on-site_Energies}
\vspace{-0.3cm}
\end{table}
Our numerical results for type $\alpha'$, $\beta'$, and $\gamma'$ polymers
(cf. Figs.~\ref{fig:EIGtypea},~\ref{fig:EIGtypeb}, and~\ref{fig:EIGtypec}),
indicate that, as, increasing $N$, a polymer is formed, the energy eigenvalues are distributed around the on-site energies of the base pairs within TB I or the bases within TB II. Hence, the HOMO (LUMO) eigenspectrum of a given polymer within TB I corresponds to the upper (lower) part of its eigenspectrum within TB II.

\begin{widetext}

\begin{figure} [h!]
\includegraphics[width=13cm]{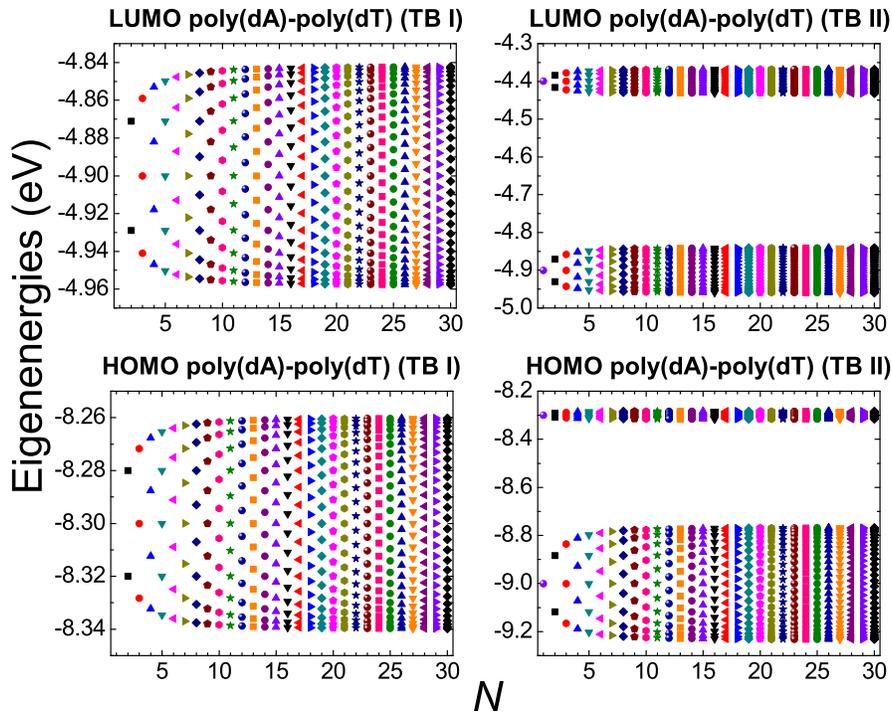}
\caption{(Color online) An example of type $\alpha'$ polymers: LUMO (first row) and HOMO (second row) eigenspectra of poly(dA)-poly(dT), for wire model (TB I, first column) and extended ladder model (TB II, second column).}
\label{fig:EIGtypea}
\end{figure}

\begin{figure}[h!]
\includegraphics[width=13cm]{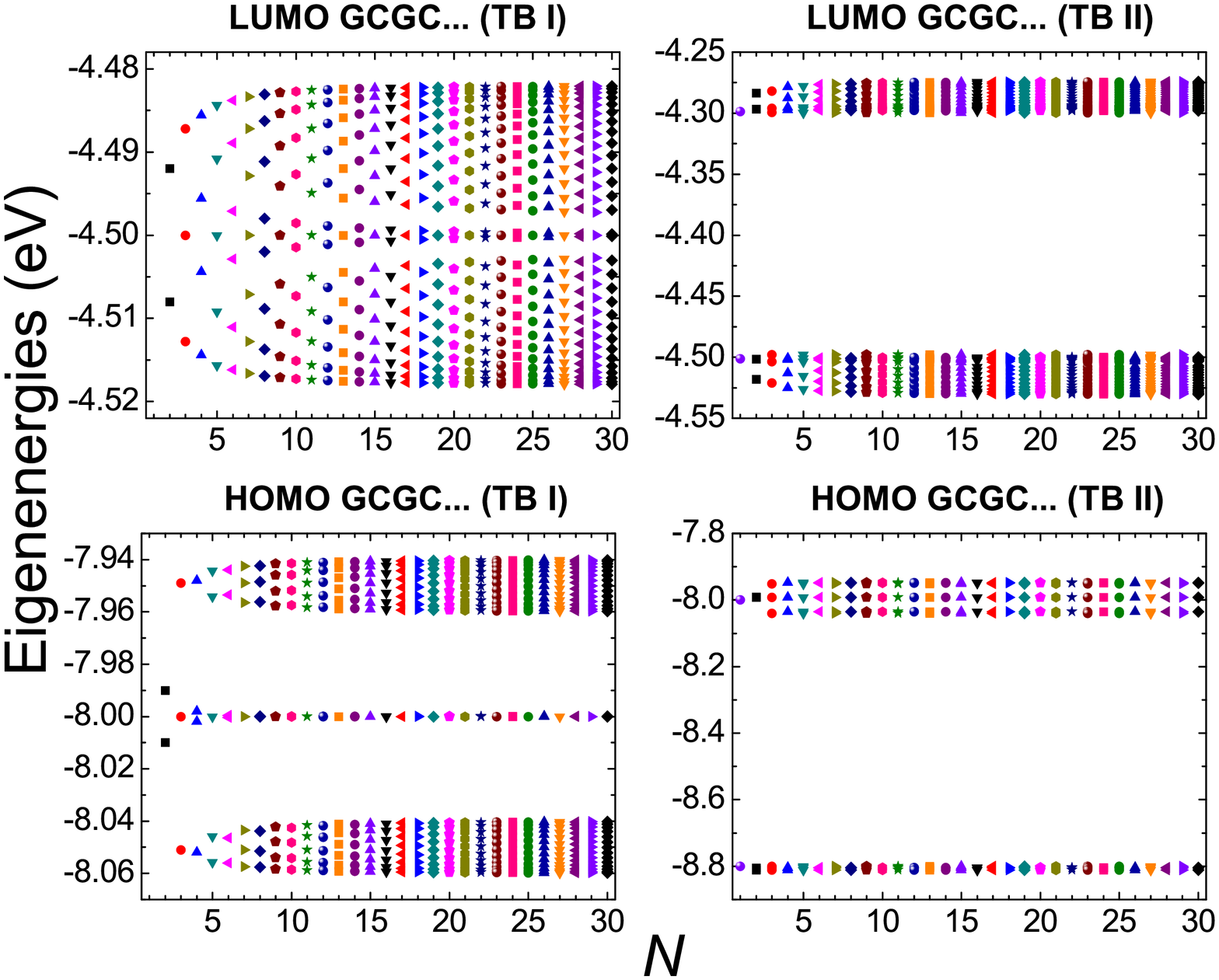}
\caption{(Color online) An example of type $\beta'$ polymers: LUMO (first row) and HOMO (second row)  eigenspectra of GCGC{\dots}, for wire model (TB I, first column) and extended ladder model (TB II, second column).
%For $N$ odd, GCGC{\dots} $\equiv$ CGCG{\dots}.
}
\label{fig:EIGtypeb}
\vspace{0.5cm}
\end{figure}

\begin{figure}[h!]
\includegraphics[width=13cm]{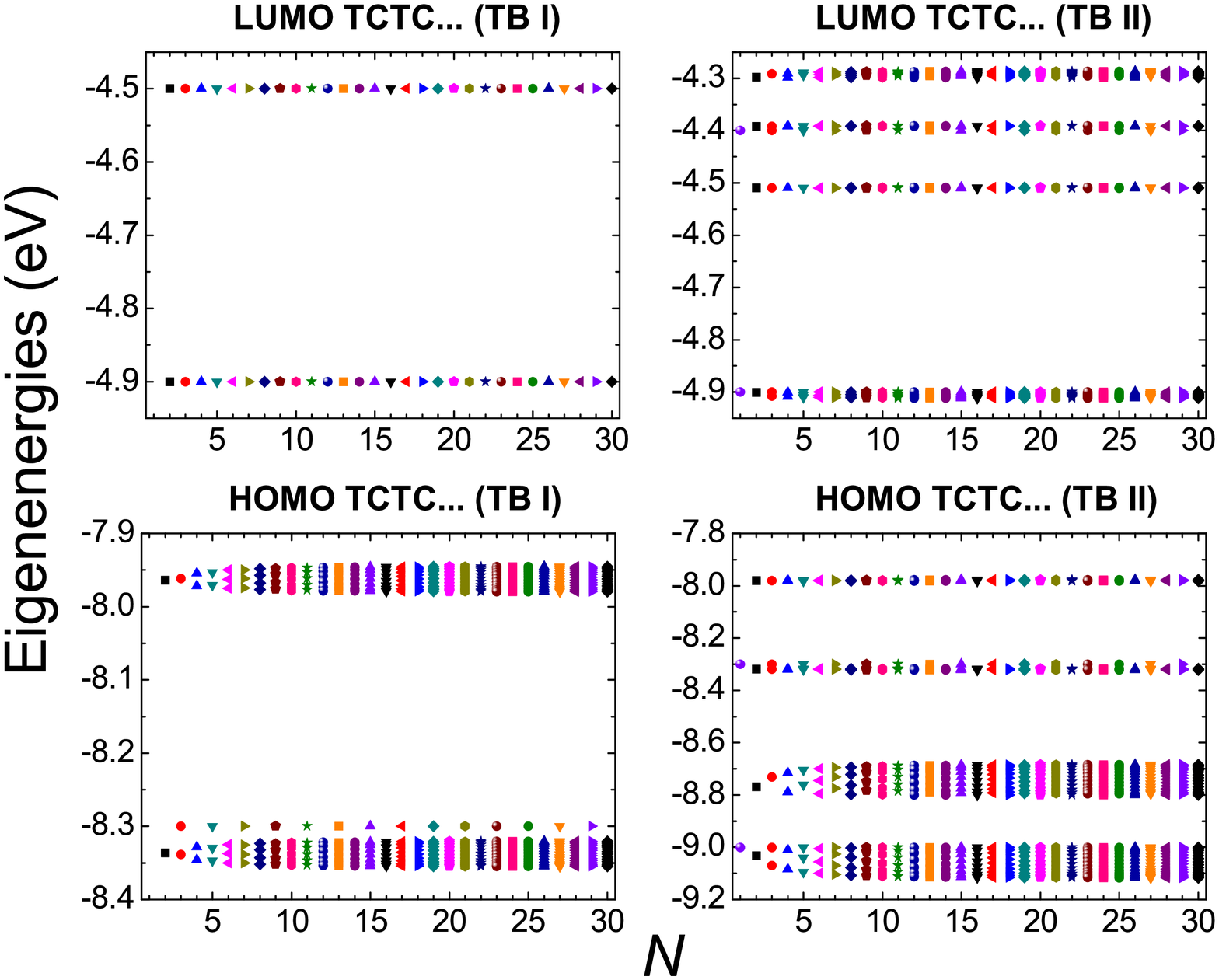}
\caption{(Color online) An example of type $\gamma'$ polymers: LUMO (first row) and HOMO (second row) eigenspectra of TCTC{\dots}, for wire model (TB I, first column) and extended ladder model (TB II, second column).
%For $N$ odd, TCTC{\dots} $\equiv$ AGAG{\dots}, while, for $N$ even, TCTC{\dots} $\equiv$ GAGA{\dots}.
}
\label{fig:EIGtypec}
\end{figure}

\end{widetext}

%%%%%%%%%%%%%%%%%%%%%%%%%%%%%%%%%%%%%%%%%%%%%%%%%%%%%%%%%%%%%%%%%%%%%%%
\subsubsection{\label{subsubsec:eigtypea} type $\alpha'$ polymers} %%%%
%%%%%%%%%%%%%%%%%%%%%%%%%%%%%%%%%%%%%%%%%%%%%%%%%%%%%%%%%%%%%%%%%%%%%%%
For TB I, an analytical expression for the eigenvalues of type $\alpha '$ polymers exists~\cite{LChMKTS:2015}. All eigenvalues are symmetric around the on-site energy $E^{bp}$ of the monomers and lie in the interval $\left[E^{bp}-2\abs{t^{bp}}, E^{bp}+2\abs{t^{bp}}\right]$. For odd $N$, the trivial eigenvalue, $E=E^{bp}$, exists.
In the left column of Fig.~\ref{fig:EIGtypea} we present the calculated HOMO/LUMO eigenspectra for an example of type $\alpha'$ polymers, poly(dA)-poly(dT).
For TB I, an analytical expression can also be found for the eigenvectors~\cite{LChMKTS:2015}. The eigenvectors (hence, the occupation probabilities, too) are \textit{eigenspectrum independent}~\cite{LChMKTS:2015}, i.e., they do not depend on the TB parameters $E^{bp}, t^{bp}$. Furthermore, the occupation probabilities display \textit{palindromicity}~\cite{LChMKTS:2015}, i.e., the occupation probability of each eigenstate of the $\mu$-th monomer is equal to that of the $(N - \mu +1)$-th monomer ($\abs{u_{\mu k}}^2 = \abs{u_{(N-\mu+1) k}}^2$).

For TB II, up to our knowledge, there are no analytical expressions for eigenvalues and eigenvectors. As an example of type $\alpha'$ polymers, we show, in the right column of Fig.~\ref{fig:EIGtypea}, the calculated HOMO/LUMO eigenspectra of poly(dA)-poly(dT).
The eigenvalues are distributed in two subbands of different width, around the on-site energies of the bases.
Furthermore, in accordance with the \textit{Observation}, the upper (lower) subband of the HOMO (LUMO) eigenspectrum corresponds to the band calculated with TB I.
For TB II, our numerical results for the eigenvectors indicate that,
for $\beta$ odd  (strand 1), $\abs{u_{\beta k}}^2 \approx \abs{u_{(2N-\beta) k}}^2$, while, for $\beta$ even (strand 2), $\abs{u_{\beta k}}^2 \approx \abs{u_{(2N-\beta + 2) k}}^2$, i.e., the occupation probabilities of the eigenstates display approximate \textit{strand-palindromicity}. For HOMO poly(dG)-poly(dC), a case where, according to the parameters used here~\cite{HKS:2010-2011}, the hopping integrals between diagonally located bases of successive monomers in the $3'$-$3'$ and $5'$-$5'$ directions are equal, strand palindomicity is strict. This also holds for all type $\alpha'$ polymers, if our extended ladder model is reduced to a simple ladder model by neglecting $3'$-$3'$ and $5'$-$5'$ inter-strand interactions.

%%%%%%%%%%%%%%%%%%%%%%%%%%%%%%%%%%%%%%%%%%%%%%%%%%%%%%%%%%%%%%%%%%%%%%
\subsubsection{\label{subsubsec:eigtypeb} type $\beta'$ polymers} %%%%
%%%%%%%%%%%%%%%%%%%%%%%%%%%%%%%%%%%%%%%%%%%%%%%%%%%%%%%%%%%%%%%%%%%%%%
As far as equivalent polymers are concerned, for $N$ even, reflection of the hamiltonian matrix $H$ on its main antidiagonal leads to the same polymers, while for $N$ odd,
GCGC{\dots}  $\equiv$  CGCG{\dots}, ATAT{\dots}  $\equiv$  TATA{\dots}.

For TB I, analytical expressions for the eigenvalues of type $\beta'$ polymers with $N$ odd exist~\cite{LChMKTS:2015}. For $N$ odd the eigenvalues can be expressed
explicitly in terms of Chebyshev zeros~\cite{Gover:1994}. All eigenvalues are symmetric around the on-site energy of the monomers, $E^{bp}$, and the trivial eigenvalue $E^{bp}$ exists. The eigenvalues lie in the interval $\left[E^{bp}-\sqrt{t_1^2+t_2^2+2\abs{t_1t_2}}, E^{bp}+\sqrt{t_1^2+t_2^2+2\abs{t_1t_2}}\right]$, where $t_1, t_2$ are the two different hopping integrals e.g., moving from the beginning to the end of the polymer, from-odd-to-even $\mu$ and from-even-to-odd $\mu$, respectively.
For $N$ even, there is no explicit formula, although a recipe to produce the eigenvalues exists~\cite{Gover:1994}.
Our numerical results show that all eigenvalues are symmetric around the on-site energy $E^{bp}$ of the monomers, and lie in the same interval as for $N$ odd.
The calculated HOMO/LUMO eigenspectrum for an example of type $\beta'$ polymers (GCGC{\dots}), displaying all the above mentioned properties,
is shown in the left column of Fig.~\ref{fig:EIGtypeb}.

For TB I and $N$ odd, analytical expressions for the eigenvectors exist~\cite{Kouachi:2006}. These eigenvectors (hence, the occupation probabilities, too) are \textit{partially eigenspectrum dependent}\cite{LChMKTS:2015}, i.e., they depend on $t_1,t_2$ but not on $E^{bp}$. Furthermore, for $\mu$ even, the occupation probability of each eigenstate of the $\mu$-th monomer is equal to that of the $(N - \mu +1)$-th monomer ($\abs{u_{\mu k}}^2 = \abs{u_{(N-\mu+1) k}}^2$), i.e., for $N$ odd, the occupation probabilities of type
$\beta'$ polymers display \textit{partial palindromicity} \cite{LChMKTS:2015}.
Finally, for $N$ odd, equivalence leads to the property $\abs{u_{\mu k}}^2(\text{YX}{\dots})= \abs{u_{(N-\mu +1)k}}^2(\text{XY}{\dots})$.
For $N$ even, we are aware of no analytical expressions for the eigenvectors, but our numerical results show that the occupation probabilities display \textit{palindromicity}~\cite{LChMKTS:2015}, i.e., for each eigenstate, the occupation probability of the $\mu$-th monomer is equal to that of the $(N - \mu +1)$-th monomer ($\abs{u_{\mu k}}^2 = \abs{u_{(N-\mu+1) k}}^2$).

For TB II, up to our knowledge, there are no analytical expressions for eigenvalues and eigenvectors. As an example of type $\beta'$ polymers, we show in the right column of Fig.~\ref{fig:EIGtypeb} the calculated HOMO/LUMO eigenspectra for GCGC{\dots}.
The eigenvalues are distributed in two subbands of different width, around the on-site energies of the bases.
Moreover, in accordance with the \textit{Observation}, the upper (lower) subband of the HOMO (LUMO) eigenspectrum corresponds to the band calculated within TB I.
For the TB II eigenvectors, for $N$ odd, equivalence leads to the property $\abs{u_{\beta k}}^2(\text{YX}{\dots})= \abs{u_{(2N-\beta +1)k}}^2(\text{XY}{\dots})$, while for $N$ even, $\abs{u_{\beta k}}^2 = \abs{u_{(2N-\beta +1)k}}^2$.
In other words, the occupation probabilities of the eigenstates display \textit{base-palindromicity}.

%%%%%%%%%%%%%%%%%%%%%%%%%%%%%%%%%%%%%%%%%%%%%%%%%%%%%%%%%%%%%%%%%%%%%%%
\subsubsection{\label{subsubsec:eigtypec} type $\gamma'$ polymers} %%%%
%%%%%%%%%%%%%%%%%%%%%%%%%%%%%%%%%%%%%%%%%%%%%%%%%%%%%%%%%%%%%%%%%%%%%%%
Here, the equivalent polymers are:
for $N$ odd,
ACAC{\dots}  $\equiv$  TGTG{\dots},
CACA{\dots}  $\equiv$  GTGT{\dots},
CTCT{\dots}  $\equiv$  GAGA{\dots},
TCTC{\dots}  $\equiv$  AGAG{\dots}, and
for $N$ even,
ACAC{\dots}  $\equiv$  GTGT{\dots},
CACA{\dots}  $\equiv$  TGTG{\dots},
CTCT{\dots}  $\equiv$  AGAG{\dots},
TCTC{\dots}  $\equiv$  GAGA{\dots}.

For TB I, analytical expressions for the eigenvalues with $N$ odd exist~\cite{LChMKTS:2015}. Let us call $E_{o(e)}^{bp}$ the on-site energy of monomers with $\mu$ odd (even),
$\Sigma=(E_o^{bp}+E_e^{bp})/2$ and $\Delta=(E_o^{bp}-E_e^{bp})/2$.
Then, the eigenvalues include $E_o^{bp}$, while the rest eigenvalues lie in the interval
$[\Sigma - \sqrt{\Delta^2 + t_1^2 + t_2^2 + 2 \abs{t_1t_2}},\allowbreak
  \Sigma + \sqrt{\Delta^2 + t_1^2 + t_2^2 + 2 \abs{t_1t_2}}]$.
For $N$ even, there is no explicit formula, although a recipe to produce the eigenvalues exists~\cite{Gover:1994}.
Our numerical results show that all eigenvalues are symmetric around $\Sigma$, and lie in the same interval as for $N$ odd.
The calculated HOMO/LUMO eigenspectrum for an example of type $\gamma'$ polymers (TCTC{\dots}), displaying all the above mentioned properties,
is shown in the left column of Fig.~\ref{fig:EIGtypec}.
For TB I and for $N$ odd, analytical expressions can also be found for the eigenvectors \cite{Alvarez:2005}. The eigenvectors (hence, the occupation probabilities, too) are \textit{eigenspectrum dependent}~\cite{LChMKTS:2015}, i.e., they depend on $E_1^{bp}, E_2^{bp}, t_1$, $t_2$. Furthermore, for $\mu$ even, the occupation probability of each eigenstate of the $\mu$-th monomer is equal to that of the $(N - \mu +1)$-th monomer ($\abs{u_{\mu k}}^2 = \abs{u_{(N-\mu+1) k}}^2$), i.e., for $N$, odd, the occupation probabilities of type $\gamma'$ polymers display \textit{partial palindromicity}.
For $N$ odd, equivalence leads to
$\abs{u_{\mu k}}^2(\text{YX}{\dots})= \abs{u_{(N-\mu +1)k}}^2 (\text{Y}_{\text{compl}}\text{X}_{\text{compl}}\dots)$.
For $N$ even, up to our knowledge, no analytical expressions for the eigenvectors exist, but equivalence leads to $\abs{u_{\mu k}}^2(\text{YX}{\dots})= \abs{u_{(N-\mu +1)k}}^2 (\text{X}_{\text{compl}}\text{Y}_{\text{compl}}\dots)$. Our numerical results show that, for all $\mu$,
$\abs{u_{\mu k}}^2(\text{YX}{\dots})= \abs{u_{\mu(N-k +1)}}^2 (\text{X}_{\text{compl}}\text{Y}_{\text{compl}}\dots)$.

For TB II, there are no analytical expressions in the literature for eigenvalues and eigenvectors, as far as we know. The calculated HOMO/LUMO eigenspectra for an example of type $\beta'$ polymers (TCTC{\dots}) are demonstrated in the right column of Fig. \ref{fig:EIGtypec}. The eigenvalues are distributed in four subbands of different width, around the on-site energies of the bases.
Moreover, in accordance with the \textit{Observation}, the two upper (lower) TB II subbands of the HOMO (LUMO) eigenspectrum correspond to the bands calculated with TB I.
For the TB II eigenvectors, equivalence leads to the properties
$\abs{u_{\beta k}}^2(\text{YX}{\dots})= \abs{u_{(2N-\beta +1)k}}^2 (\text{Y}_{\text{compl}}\text{X}_{\text{compl}}\dots)$, for $N$ odd, and $\abs{u_{\beta k}}^2(\text{YX}{\dots})= \abs{u_{(2N-\beta +1)k}}^2 (\text{X}_{\text{compl}}\text{Y}_{\text{compl}}\dots)$, for $N$ even. Our numerical results indicate that there are no palindromic properties.

%%%%%%%%%%%%%%%%%%%%%%%%%%%%%%%%%%%%%%%%%%%%%%%%%%%%
\subsection{\label{subsec:DOS} Density of States} %%
\begin{figure}[h!]
\includegraphics[width=7.5cm]{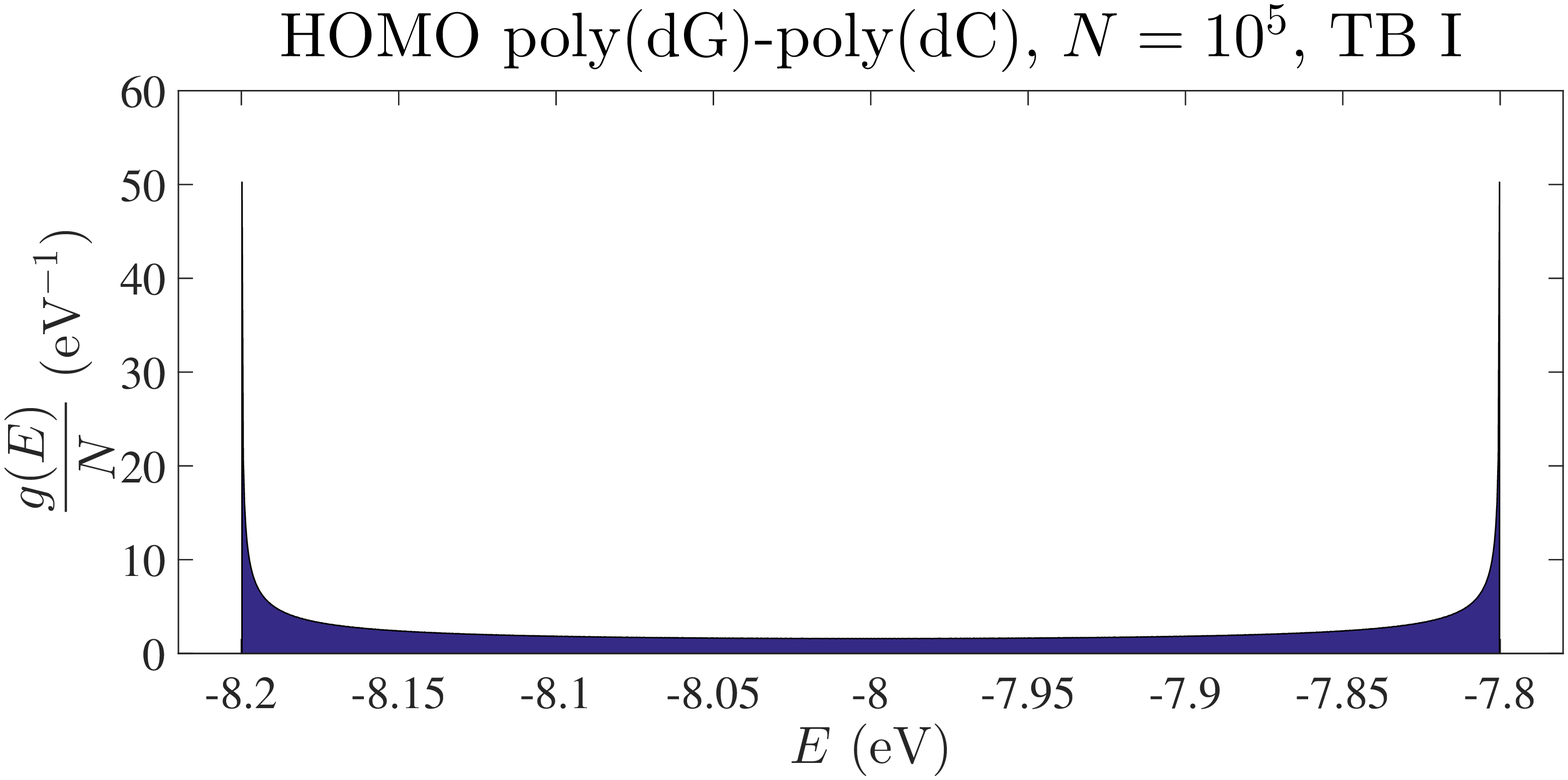}
\includegraphics[width=7.5cm]{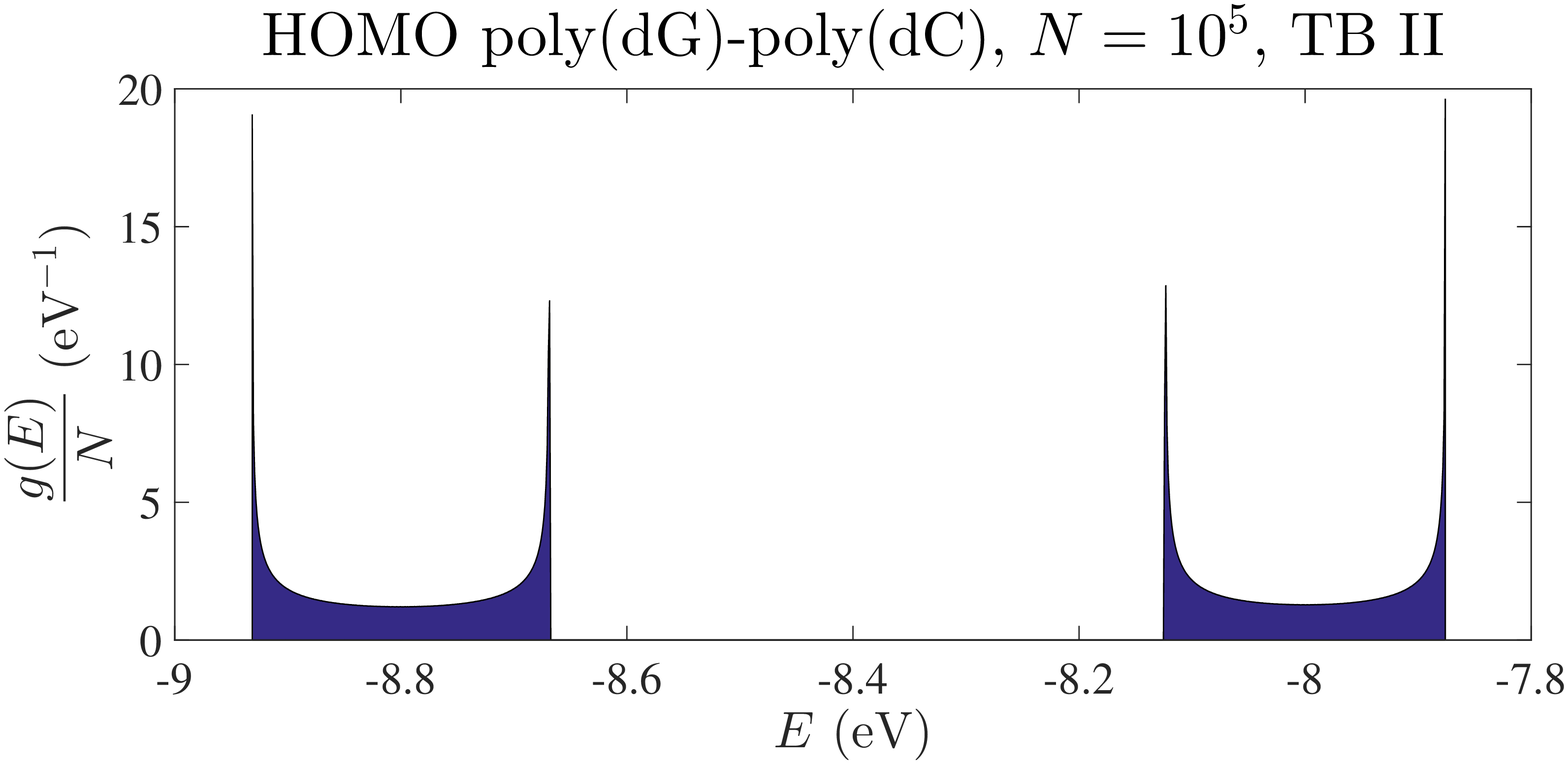}
\caption{(Color online) DOS for an example of type $\alpha'$ polymers, poly(dG)-poly(dC) ($N = 10^5$, HOMO), for the base-pair (TB I, top) and the single-base (TB II, bottom) approaches.}
\label{fig:DOStypea}
\end{figure}

\begin{figure}[h!]
\includegraphics[width=7.5cm]{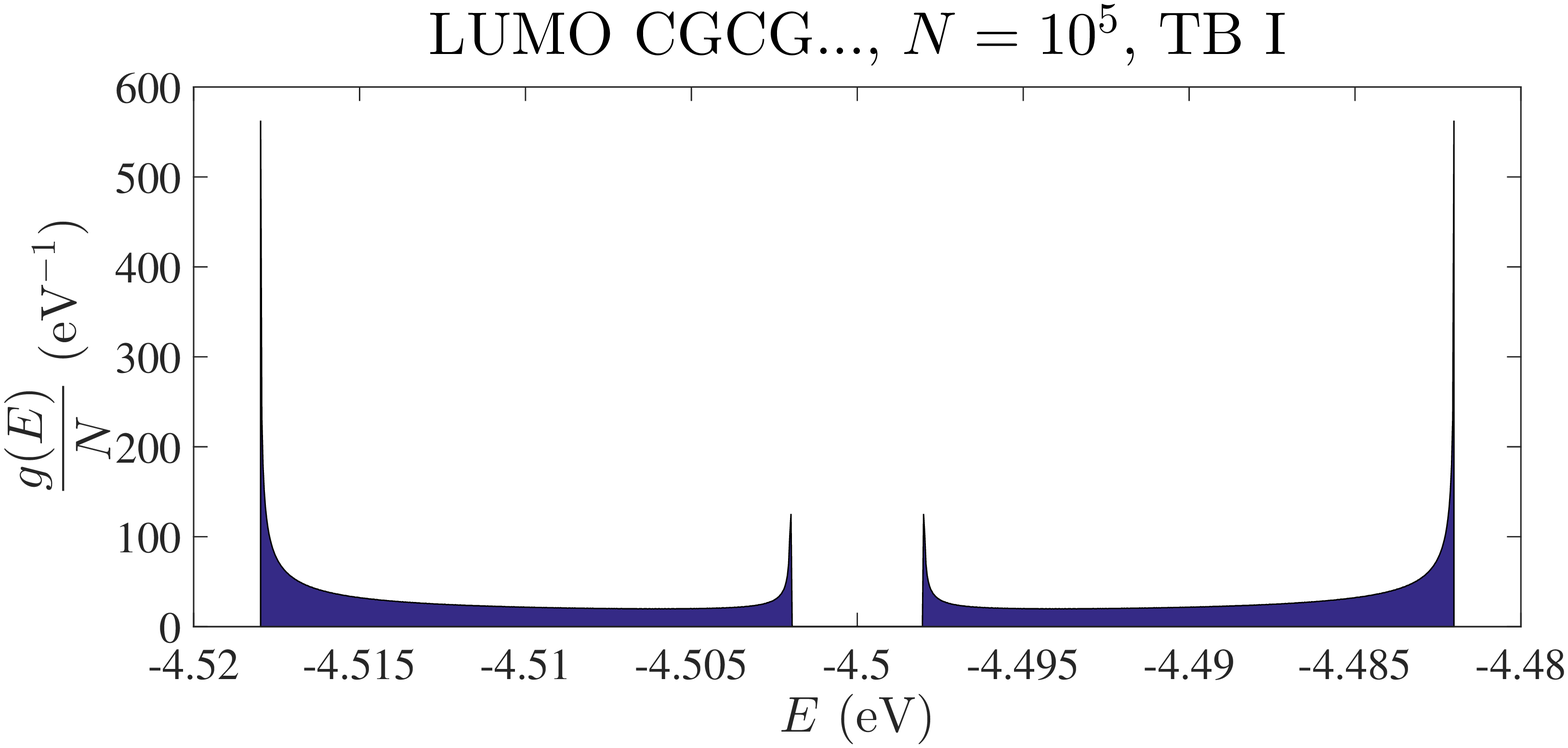}
\includegraphics[width=7.5cm]{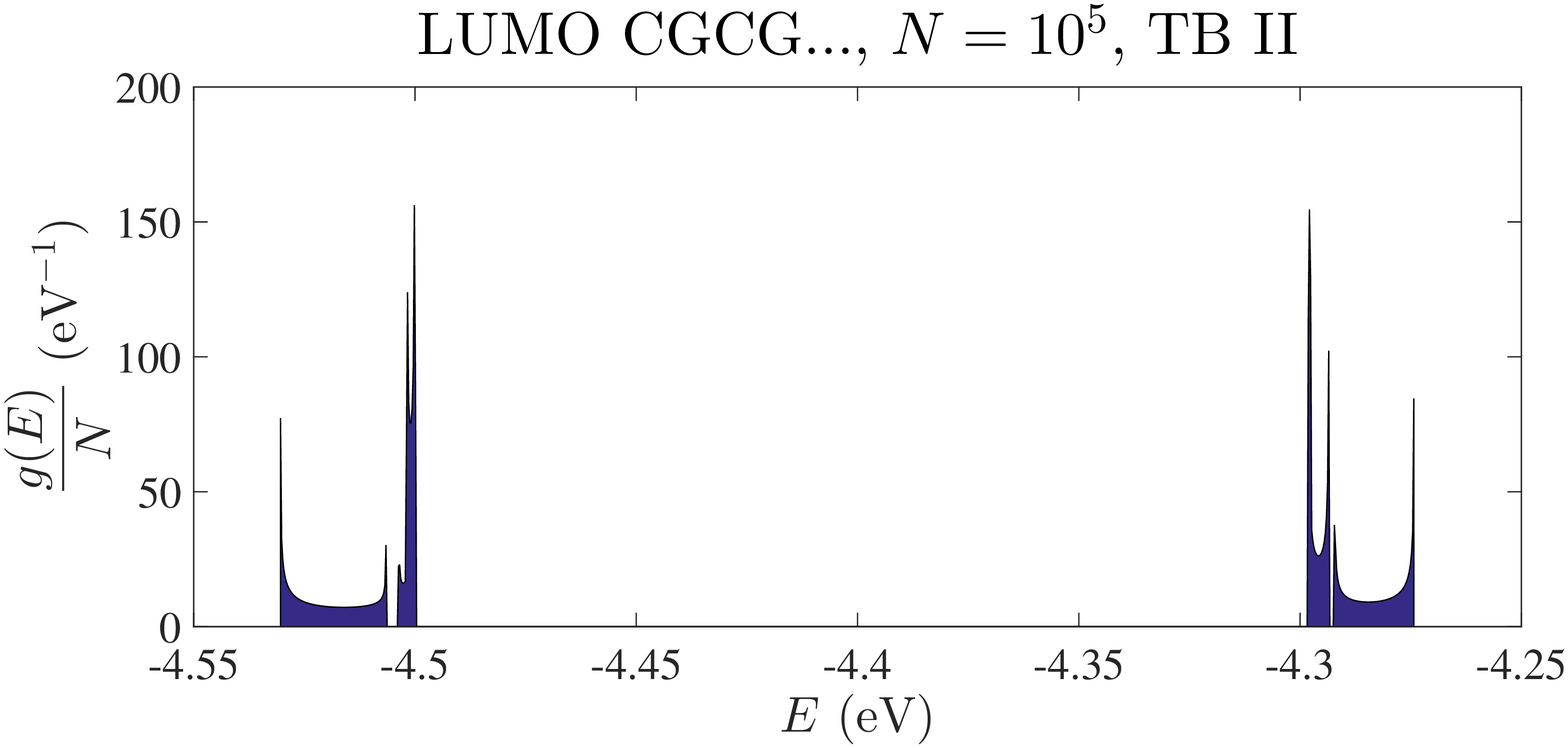}
\caption{(Color online) DOS for an example of type $\beta'$ polymers, CGCG{\dots} ($N = 10^5$, LUMO), for the base-pair (TB I, top) and the single-base (TB II, bottom) approaches. }
\label{fig:DOStypeb}
\end{figure}

\begin{figure}[h!]
\includegraphics[width=7.5cm]{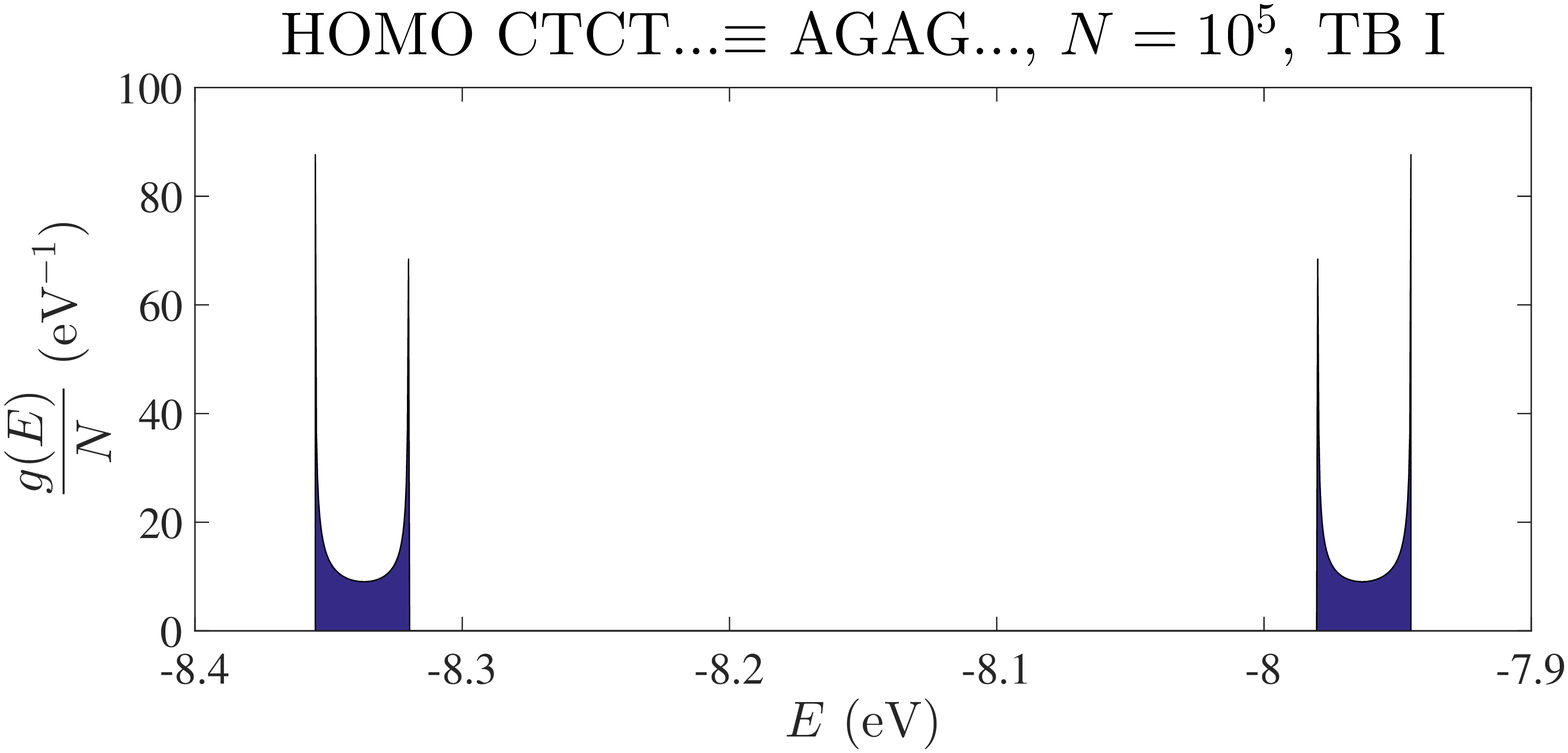}
\includegraphics[width=7.5cm]{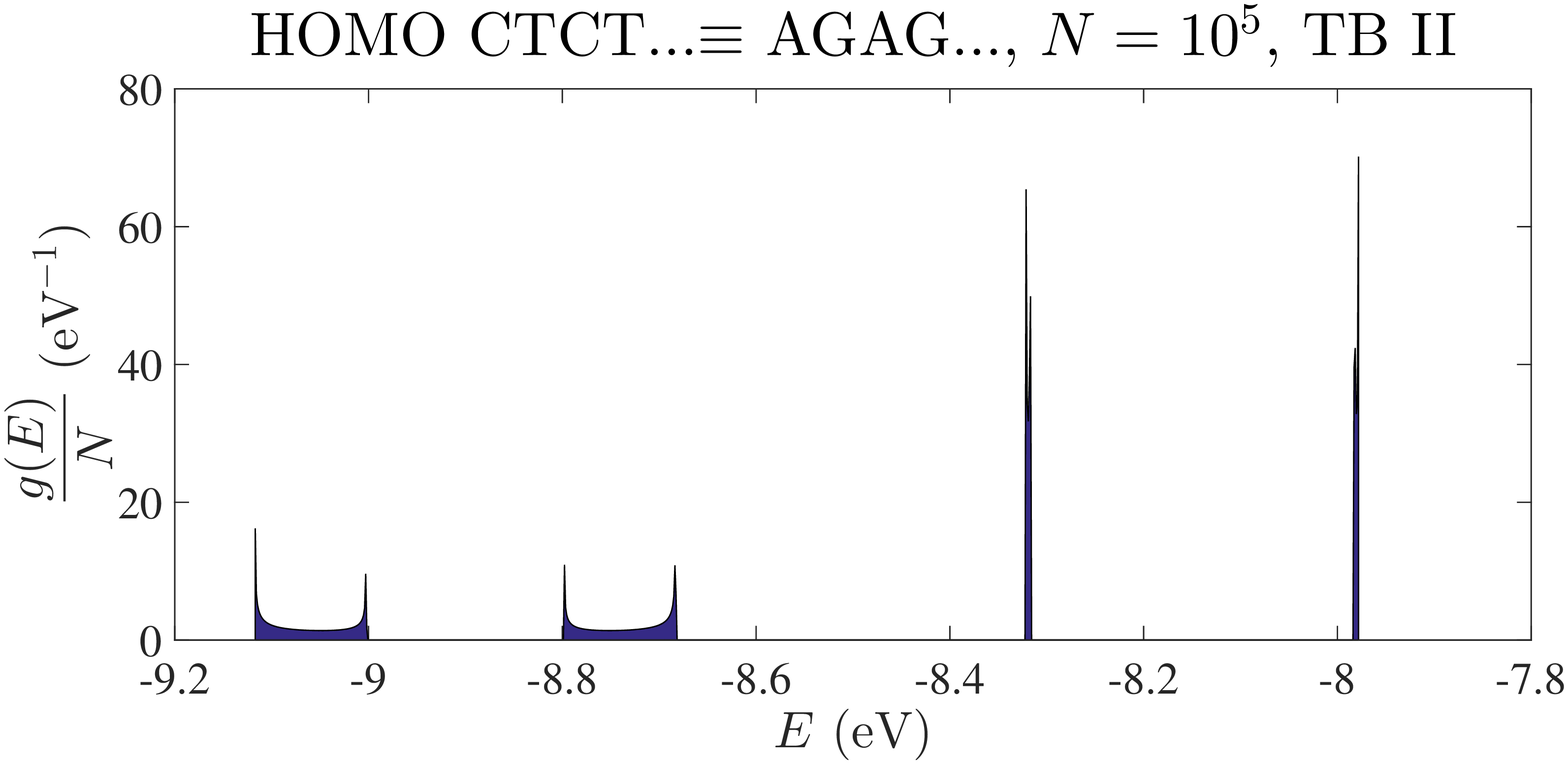}
\caption{(Color online) DOS for an example of type $\gamma'$ polymers, CTCT{\dots} $\equiv$ AGAG{\dots} ($N = 10^5$, HOMO) for the base-pair (TB I, top) and the single-base (TB II, bottom) approaches.}
\label{fig:DOStypec}
\end{figure}

%%%%%%%%%%%%%%%%%%%%%%%%%%%%%%%%%%%%%%%%%%%%%%%%%%%%
For TB I or TB II, the DOS can be determined directly by the eigenspectra (cf. Eq.~\eqref{DOS}). It represents nicely the corresponding eigenspectral  properties (cf. \S~\ref{subsec:eigenSpectraVectors}). In Figs.~\ref{fig:DOStypea},~\ref{fig:DOStypeb},~and~\ref{fig:DOStypec}, we illustrate the numerically determined DOS for some representative examples of type $\alpha'$, $\beta'$, and $\gamma'$ polymers, respectively, for $N = 10^5$, i.e., in the large $N$ limit. We observe that, due to the fact that the eigenenergies become denser and denser as we approach the band or subband edges,   van Hove Singularities (vHS) occur at the edges of each band or subband. We also notice that, in the large $N$ limit, the polymer boundaries play insignificant role in the electronic structure, hence, for the same set of TB parameters, the polymers' DOS is essentially the same. For example, in the large $N$ limit, either GCGC... or CGCG..., either $N$ odd or $N$ even have practically the same DOS.
In some simpler cases, the DOS can be analytically obtained.
For example, for type $\alpha'$ polymers, within TB I, in the large $N$ limit and using periodic boundaries, i.e., for \textit{cyclic} type $\alpha'$ polymers~\cite{LChMKTS:2015},
\begin{equation} \label{DOStypeaTBI}
g(E)= \frac{N}{\pi \sqrt{4(t^{bp})^2 - (E-E^{bp})^2}}.
\end{equation}

For TB I, the numerically derived DOS for type $\alpha'$ polymers (cf. Fig.~\ref{fig:DOStypea}) is in accordance with Eq.~\eqref{DOStypeaTBI}, because in the large $N$ limit the boundary conditions play insignificant role.
In Fig.~\ref{fig:DOStypea}, for TB I, there is no minigap, but
for TB II there is a minigap of $\approx$ 0.545 eV; in accordance with the \textit{Observation}, the upper subband of the HOMO band calculated with TB II, corresponds to the HOMO band calculated with TB I. The minigap is mainly due to the different HOMO on-site energies of the two bases ($-8.0$ eV for G, $-8.8$ eV for C)~\cite{HKS:2010-2011}.

A DOS example in type $\beta'$ polymers is shown in Fig.~\ref{fig:DOStypeb}.
For TB I, there is a small ($\approx$ \emph{0.004 eV}) minigap.
For TB II, there is a minigap of $\approx$ 0.200 eV; in accordance with the \textit{Observation}, the lower subband of the LUMO band calculated with TB II corresponds to the LUMO band calculated with TB I.
For TB II, there are two additional small ($\approx$ \emph{0.003 eV}, 0.001 eV) minigaps, hardly noticeable at this scale.
The \emph{underlined} TB II minigap corresponds to the TB I minigap, also \emph{underlined}.
For TB II, apart from the vHSs at the subband edges, there is one vHS inside the second subband, hardly seen at this scale.

A DOS example in type $\gamma'$ polymers is shown in Fig.~\ref{fig:DOStypec}.
For TB I, there is a minigap a little greater than \emph{0.340 eV}, mainly due to the different HOMO on-site energies of the two base pairs ($-8.0$ eV for G-C, $-8.3$ eV for A-T)~\cite{HKS:2010-2011}.
For TB II, four minibands are formed approximately around the HOMO on-site energies of the four bases
($-9.0$ eV for T, $-8.8$ eV for C, $-8.3$ eV for A and $-8.0$ eV for G)~\cite{HKS:2010-2011},
with three relevant minigaps (0.205 eV,  0.362 eV, \emph{0.334 eV}).
Two of these minibands are very narrow. In accordance with the \textit{Observation}, the higher two subbands of the HOMO band calculated with TB II correspond to the HOMO band calculated with TB I.
The \emph{underlined} TB II minigap corresponds to the TB I minigap, also \emph{underlined}.

%\end{widetext}

%%%%%%%%%%%%%%%%%%%%%%%%%%%%%%%%%%%%%%%%%%%%%%%%%%%%
\subsection{\label{subsec:HLGaps} HOMO-LUMO gaps} %%
\begin{figure}[h!]
\includegraphics[width=8.5cm]{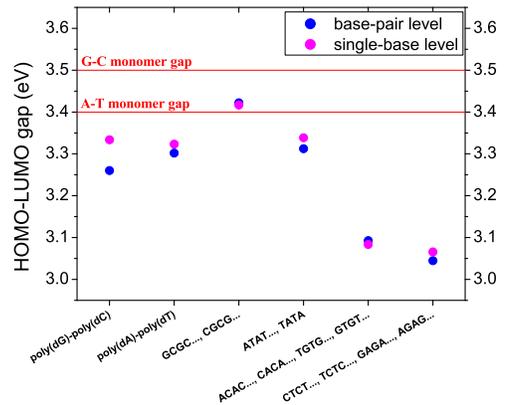}
\caption{(Color online) HOMO-LUMO gaps of type $\alpha'$, $\beta'$, and $\gamma'$ polymers, for the base-pair (TB I, blue dots) and the single-base (TB II, purple dots) approaches. The red lines denote the HOMO-LUMO gaps of the two possible monomers.}
\label{fig:gaps}
\end{figure}

%%%%%%%%%%%%%%%%%%%%%%%%%%%%%%%%%%%%%%%%%%%%%%%%%%%%
In Fig.~\ref{fig:gaps}, we present the HOMO-LUMO energy gaps, in the large $N$ limit, for all types of polymers.
Both TB approaches predict similar gaps, in the range $\approx 3.04$ - $3.42$ eV.
For TB I, the HOMO-LUMO gaps can also be derived analytically, from the maxima and minima of the HOMO and LUMO eigenspectra, respectively (cf. \S~\ref{subsec:eigenSpectraVectors}). We also compare the polymer gaps with the two possible monomer gaps. The decrease of the energy gap, as we move from monomer to polymer, is larger for type $\gamma'$ polymers.

%\pagebreak

%%%%%%%%%%%%%%%%%%%%%%%%%%%%%%%%%%%%%%%%%%%%%%%%%%%%%%%%%%%%%%%%%%%%%%%
\subsection{\label{subsec:MeanProbs} Mean over time probabilities} %%%%
%%%%%%%%%%%%%%%%%%%%%%%%%%%%%%%%%%%%%%%%%%%%%%%%%%%%%%%%%%%%%%%%%%%%%%%
Within TB I, from Eq.~\eqref{probabilities} and the \textit{initial condition} (carrier initially placed at the first monomer),
it follows that the mean over time probability to find the extra carrier at the $\mu$-th monomer is
\begin{equation} \label{meanProbbp}
\expval{\abs{C_\mu(t)}^2} = \sum_{k=1}^{N} v_{1k}^2 v_{\mu k}^2.
\end{equation}
Within TB II, from Eq.~\eqref{probabilities} and the \textit{initial condition 1} (carrier initially placed at the first base) or the \textit{initial condition 2} (carrier initially placed at the second base),
it follows that the mean probability to find the extra carrier at the $\beta$-th base is
\begin{equation} \label{meanProbsb}
\expval{\abs{C_\beta(t)}^2} =
\begin{cases}\displaystyle
\sum_{k=1}^{2N} v_{1k}^2 v_{\beta k}^2 \\ \displaystyle \sum_{k=1}^{2N} v_{2k}^2 v_{\beta k}^2
\end{cases}.
\end{equation}
From Eqs.~\eqref{meanProbbp} and \eqref{meanProbsb}, we conclude that the \textit{palindromicity} and \textit{eigenspectrum (in)dependence} properties for the occupation probabilities, presented in \S~\ref{subsec:eigenSpectraVectors}, hold also for the mean over time probabilities.
Finally, for equivalent polymers it can be shown that
in TB I $\langle \abs{C_N(t)}^2 \rangle_{\text{YX\dots}} = \langle \abs{C_N(t)}^2 \rangle_{\text{equiv(YX\dots)}}$,
while in TB II $\langle \abs{C_{2N}(t)}^2 \rangle_{\text{YX\dots}} = \langle \abs{C_{2N}(t)}^2 \rangle_{\text{equiv(YXYX\dots)}}$ (for initial condition 1) and
$\langle \abs{C_{2N-1}(t)}^2 \rangle_{\text{YXYX\dots}} = \langle \abs{C_{2N-1}(t)}^2 \rangle_{\text{equiv(YXYX\dots)}}$
(for initial condition 2).

%%%%%%%%%%%%%%%%%%%%%%%%%%%%%%%%%%%%%%%%%%%%%%%%%%%%%%%%%%%%%%%%%%%%%%%

\begin{figure} [h!]
\centering
\includegraphics[width=0.5\textwidth]{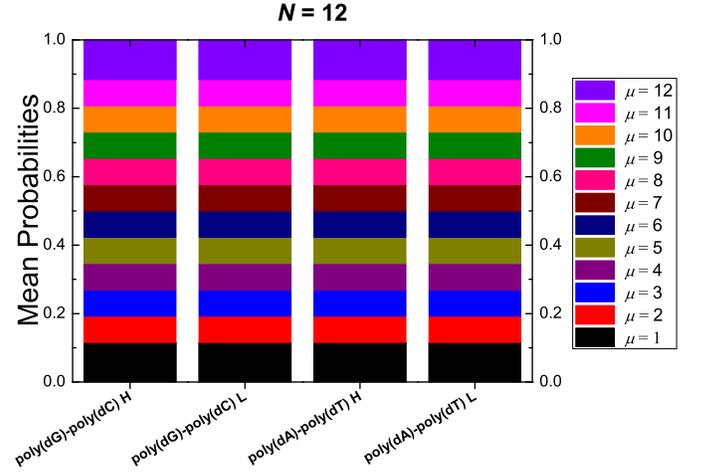}
\caption{(Color online) Type $\alpha'$ polymers.
TB I and \textit{initial condition} (extra carrier initially at the first base pair).
Mean over time probabilities to find an extra hole (HOMO) or electron (LUMO) at each base pair.
Here $N=12$.}
\label{fig:meanprobstypea-bp}
\end{figure}

\begin{widetext}

\begin{figure}[h!]
\centering
\includegraphics[width=\textwidth]{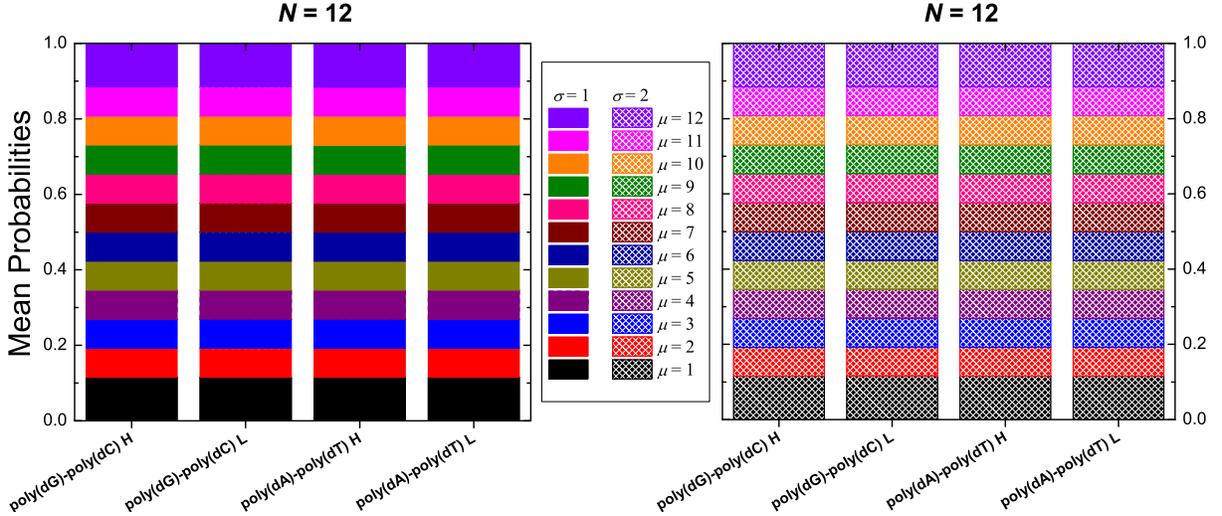}
\caption{(Color online) Type $\alpha'$ polymers.
TB II and either \textit{initial condition 1} (left) or \textit{initial condition 2} (right), i.e.,
extra carrier initially at the first or the second base of the first base pair, respectively.
Mean over time probabilities to find an extra hole (HOMO) or electron (LUMO) at each base.
Here $N=12$.}
\label{fig:meanprobstypea-sb}
\end{figure}

\end{widetext}

%%%%%%%%%%%%%%%%%%%%%%%%%%%%%%%%%%%%%%%%%%%%%%%%%%%%%%%%%%%%%%%%%%%%%%%%%%%
\subsubsection{\label{subsubsec:meanProbstypea} type $\alpha'$ polymers} %%
%%%%%%%%%%%%%%%%%%%%%%%%%%%%%%%%%%%%%%%%%%%%%%%%%%%%%%%%%%%%%%%%%%%%%%%%%%%
In Figs.~\ref{fig:meanprobstypea-bp}-\ref{fig:meanprobstypea-sb},
we show an example (for $N=12$) of our numerical results for the mean over time probabilities
to find an extra hole or electron at
(I) each base pair according to TB I and the \textit{initial condition} (Fig.~\ref{fig:meanprobstypea-bp}), and
(II) each base according to TB II and the \textit{initial condition 1} or the \textit{initial condition 2}
(Fig.~\ref{fig:meanprobstypea-sb}), for type $\alpha'$ polymers.

For TB I, the mean over time probabilities to find the carrier at a specific monomer display \textit{palindromicity} and \textit{eigenspectrum independence}~\cite{LChMKTS:2015}. Specifically, it can analytically be shown that
\begin{subequations} \label{meanProbtypeabp}
\begin{equation}
\expval{\abs{C_1(t)}^2} = \expval{\abs{C_N(t)}^2} = \frac{3}{2(N+1)}, \forall N \geq 2,
\end{equation}
\begin{equation}
\expval{\abs{C_2(t)}^2} = \dots = \expval{\abs{C_{N-1}(t)}^2} = \frac{1}{N+1}, \forall N \geq 3.
\end{equation}
\end{subequations}

For TB II, the mean over time probabilities to find the carrier at a specific base display \textit{approximate strand-palindromicity}. Moreover, adding the mean probabilities of the bases that constitute each monomer, it follows that the mean probabilities to find the carrier at a specific monomer are approximately palindromes and almost identically equal, for all cases, to the mean probabilities within TB I, which are strictly palindromes (cf. Eq.~\eqref{meanProbtypeabp}).
This quantitative agreement suggests that the eigenspectrum independence predicted within the simpler TB I approach, leads to essentially the same results as those derived by the more complicated TB II approach.
In Fig.~\ref{fig:meanprobstypea-sb} we observe that, within TB II, the carrier moves almost exclusively through the strand it was initially placed at, i.e. carrier movement is mainly of intra-strand character. Furthermore, within TB II, our results for the two initial conditions are in complete agreement.

%\pagebreak

%%%%%%%%%%%%%%%%%%%%%%%%%%%%%%%%%%%%%%%%%%%%%%%%%%%%%%%%%%%%%%%%%%%%%%%%%%
\subsubsection{\label{subsubsec:meanProbstypeb} type $\beta'$ polymers} %%
%%%%%%%%%%%%%%%%%%%%%%%%%%%%%%%%%%%%%%%%%%%%%%%%%%%%%%%%%%%%%%%%%%%%%%%%%%
In Figs.~\ref{fig:meanprobstypeb-bp}-\ref{fig:meanprobstypeb-sb}
we present examples of our numerical results for type $\beta'$ polymers, for
the mean over time probabilities to find an extra hole or electron
(I) at each base pair according to TB I and the \textit{initial condition} (Fig.~\ref{fig:meanprobstypeb-bp}), and
(II) at each base according to TB II and the \textit{initial condition 1} or the \textit{initial condition 2}
(Fig.~\ref{fig:meanprobstypeb-sb}).

For TB I, the mean probabilities to find the carrier at a specific monomer display~\cite{LChMKTS:2015}
\textit{partial eigenspectrum dependence} (i.e., dependence on the hopping parameters but not on the on-site energy),
\textit{partial palindromicity} (i.e., only for even $\mu$) for $N$ odd and \textit{palindromicity} (i.e., for all $\mu$) for $N$ even.

For TB II, for $N$ even, the mean probabilities to find the carrier at a specific base display
\textit{base-palindromicity}. Moreover, adding the mean probabilities of the bases that constitute each base pair,
the mean probabilities to find the carrier at a specific base pair are palindromes, in accordance with the prediction of TB I.
In Fig.~\ref{fig:meanprobstypeb-sb}, we observe that within TB II, the carrier moves preferably through the bases that are identical with the one it was initially placed at, in other words it moves crosswise through identical bases, i.e., carrier movement is mainly of inter-strand character.

For $N$ odd, both TB approaches show that there are some cases, in which the carrier hardly moves from its initial site.
If we add or subtract a monomer, i.e. for $N$ even, both TB approaches show that a large percentage of the carrier is transferred at the end monomer. Furthermore, both TB approaches show that the mean probability to find the carrier at the last monomer is generally bigger for $N$ even than for $N$ odd. \\

%%%%%%%%%%%%%%%%%%%%%%%%%%%%%%%%%%%%%%%%%%%%%%%%%%%%%%%%%%%%%%%%%%%%%%%%%%%
\subsubsection{\label{subsubsec:meanProbstypec} type $\gamma'$ polymers} %%
%%%%%%%%%%%%%%%%%%%%%%%%%%%%%%%%%%%%%%%%%%%%%%%%%%%%%%%%%%%%%%%%%%%%%%%%%%%
In Figs.~\ref{fig:meanprobstypec-bp}~and~\ref{fig:meanprobstypec-sb},
we present examples of our numerical results for
the mean over time probabilities to find an extra hole or electron
(I) at each base pair according to TB I and the \textit{initial condition} (Fig.~\ref{fig:meanprobstypec-bp}), and
(II) at each base according to TB II and the \textit{initial condition 1} or the \textit{initial condition 2}
(Fig.~\ref{fig:meanprobstypec-sb}), for type $\gamma'$ polymers.

In TB I, given that
$\abs{u_{\mu k}}^2(\text{YX}{\dots}) = \abs{u_{\mu(N-k +1)}}^2 (\text{X}_{\text{compl}}\text{Y}_{\text{compl}}\dots)$,
for all $\mu$, Eq.~\eqref{meanProbbp} leads to identical mean probabilities for
(\textit{i})   TCTC{\dots} and GAGA{\dots},
(\textit{ii})  CTCT{\dots} and AGAG{\dots},
(\textit{iii}) ACAC{\dots} and GTGT{\dots}, and
(\textit{iv})  CACA{\dots} and TGTG{\dots}.

In Fig. \ref{fig:meanprobstypec-bp} we observe that within TB I, the carrier moves preferably through the monomers that are identical with the one it was initially placed at, i.e., from the first monomer to the third, and so forth.
Within TB II, the carrier moves preferably through the bases that are identical with the one it was initially placed at, i.e.,
it moves through the same strand from the one or the other base of the first monomer to the identical base of the third monomer, and so forth, i.e., carrier movement is mainly of inter-strand character.

Both TB approaches show that the mean probability to find the carrier at the last monomer is bigger for $N$ odd than for $N$ even, cf. Figs.~\ref{fig:meanprobstypec-bp}-\ref{fig:meanprobstypec-sb}.

\begin{widetext}

\begin{figure} [h!]
\centering
\includegraphics[width=0.45\textwidth]{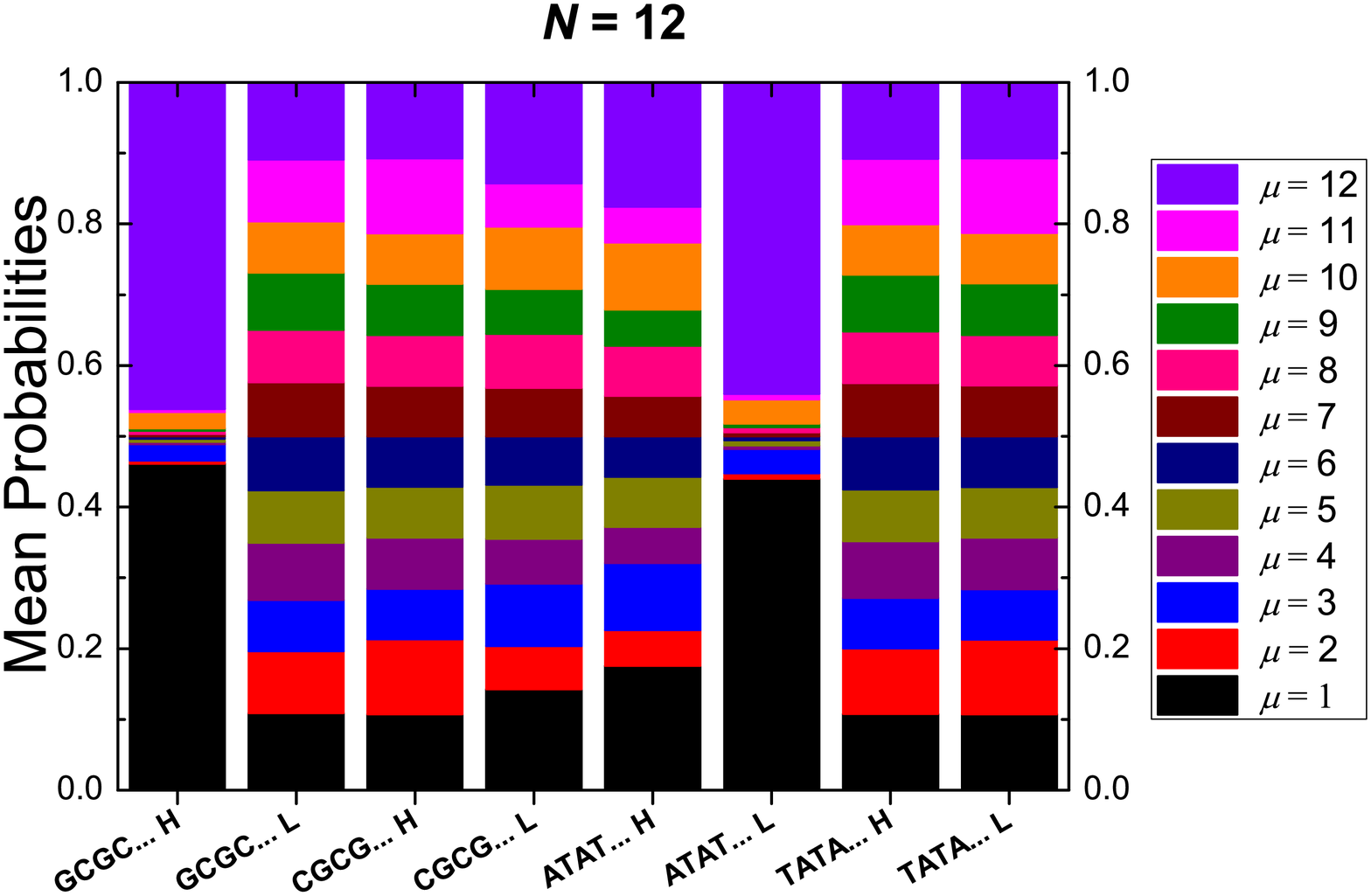}
\includegraphics[width=0.45\textwidth]{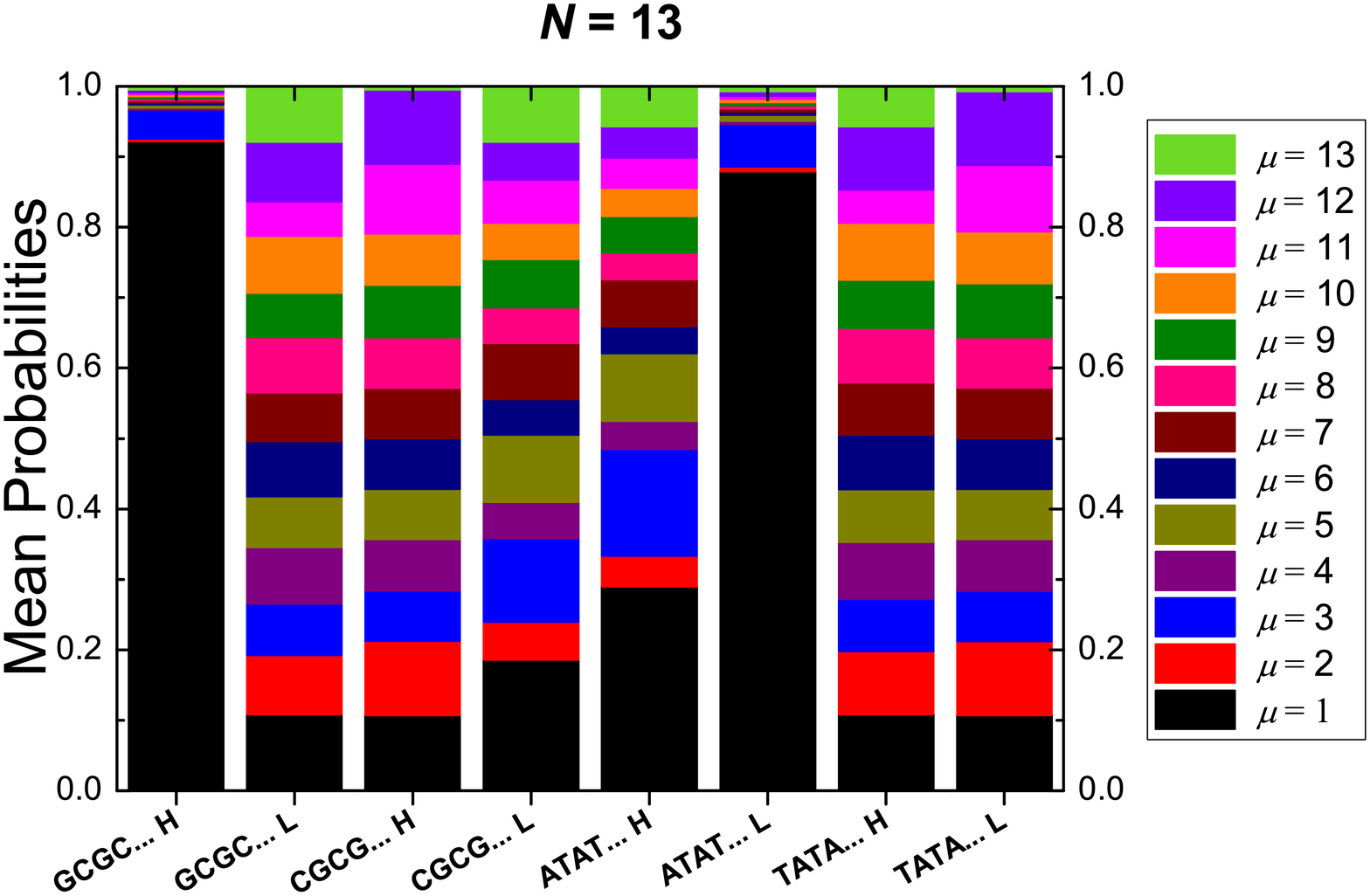}
\caption{(Color online) Type $\beta'$ polymers.
TB I and \textit{initial condition} (extra carrier initially at the first base pair).
Mean over time probabilities to find an extra hole (HOMO) or electron (LUMO) at each base pair.
$N$ even (here $N=12$, left) or $N$ odd (here, $N=13$, right).}
\label{fig:meanprobstypeb-bp}
\end{figure}

\begin{figure}[h!]
\centering
\includegraphics[width=\textwidth]{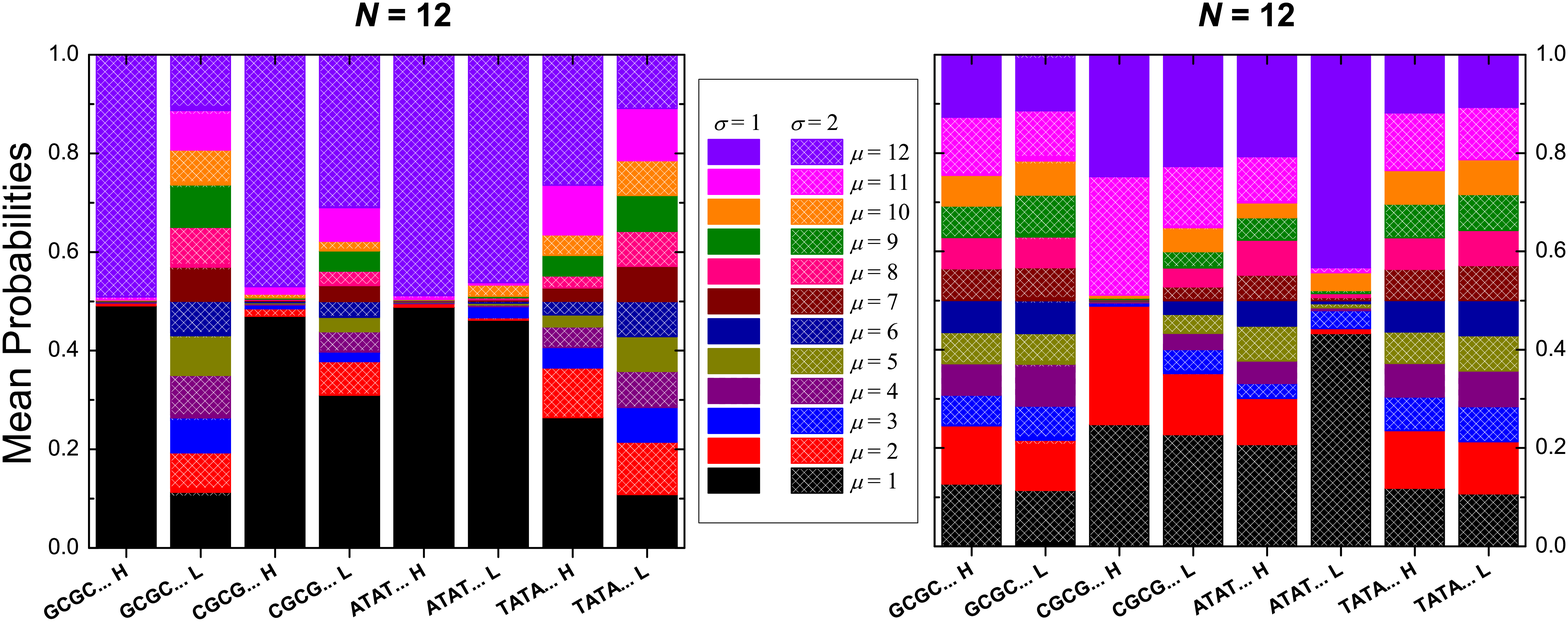}
\includegraphics[width=\textwidth]{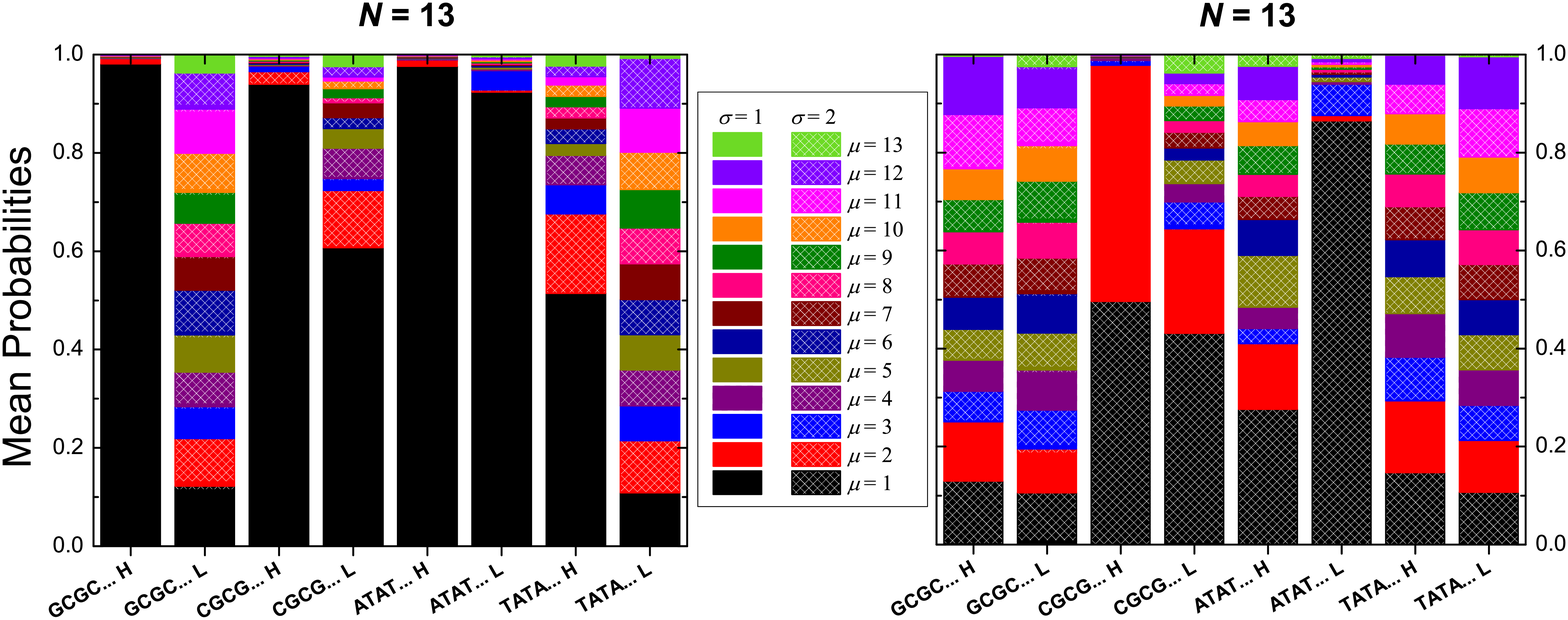}
\caption{(Color online) Type $\beta'$ polymers.
TB II and either \textit{initial condition 1} (left) or \textit{initial condition 2} (right), i.e.,
extra carrier initially at the first or the second base of the first base pair, respectively.
Mean over time probabilities to find an extra hole (HOMO) or electron (LUMO) at each base.
$N$ even (upper panels, here $N=12$) or $N$ odd (lower panels, here $N=13$).}
\label{fig:meanprobstypeb-sb}
\end{figure}

\begin{figure}[h!]
\centering
\includegraphics[width=0.45\textwidth]{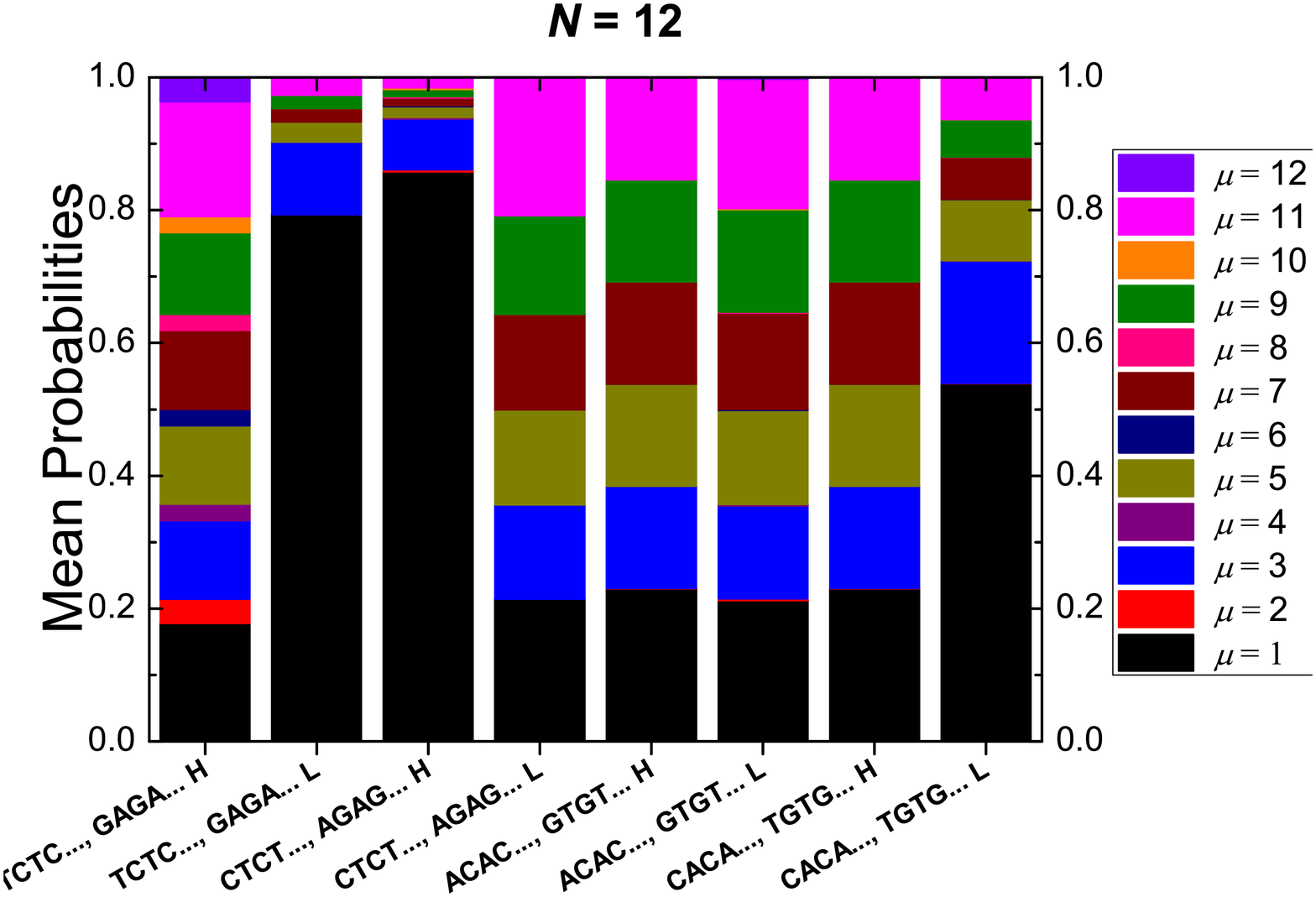}
\includegraphics[width=0.45\textwidth]{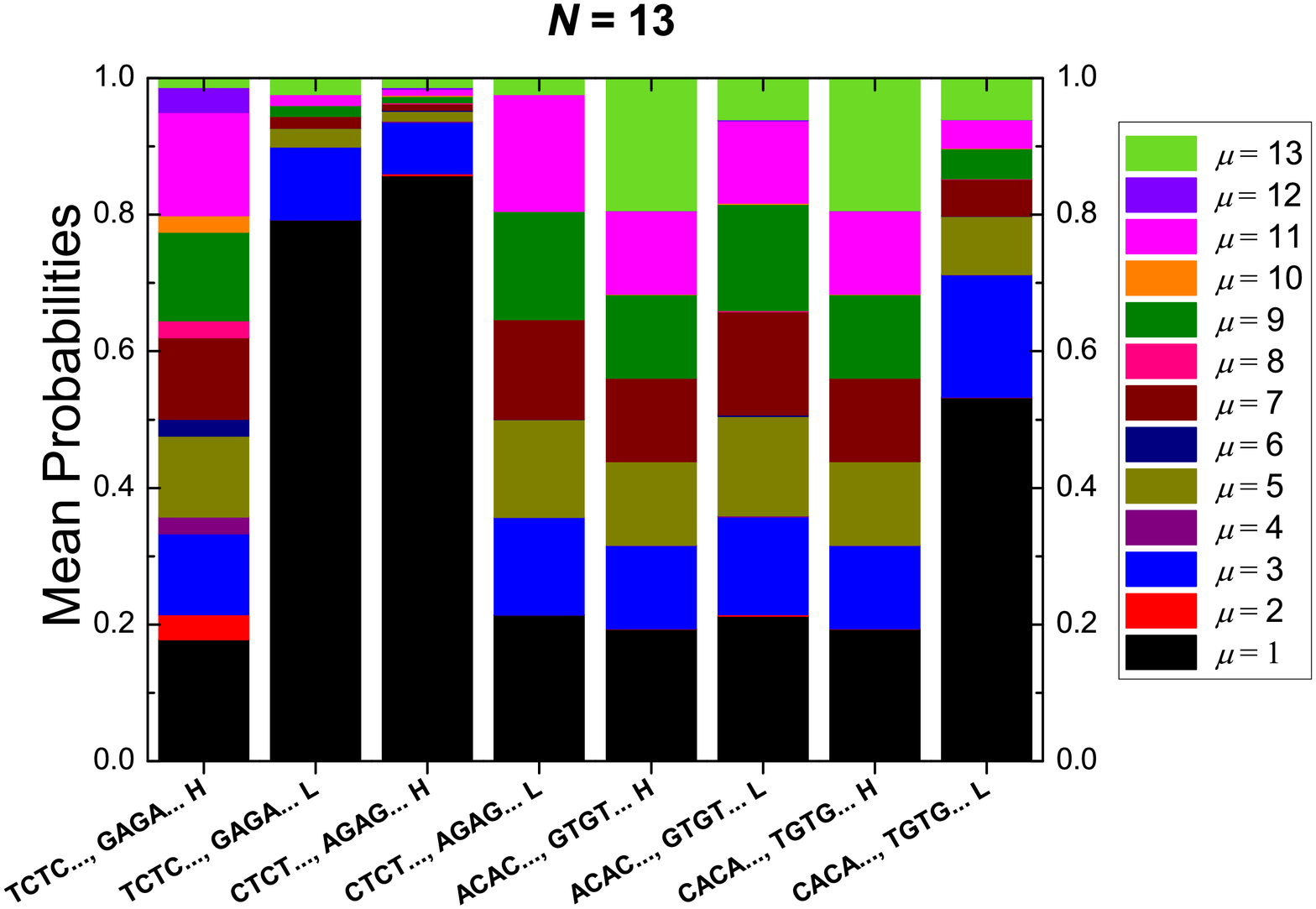}
\caption{(Color online) Type $\gamma'$ polymers.
TB I and \textit{initial condition} (extra carrier initially at the first base pair).
Mean over time probabilities to find an extra hole (HOMO) or electron (LUMO) at each base pair.
$N$ even (here $N=12$, left) or $N$ odd (here $N=13$, right).}
\label{fig:meanprobstypec-bp}
\end{figure}

\begin{figure}[h!]
\centering
\includegraphics[width=\textwidth]{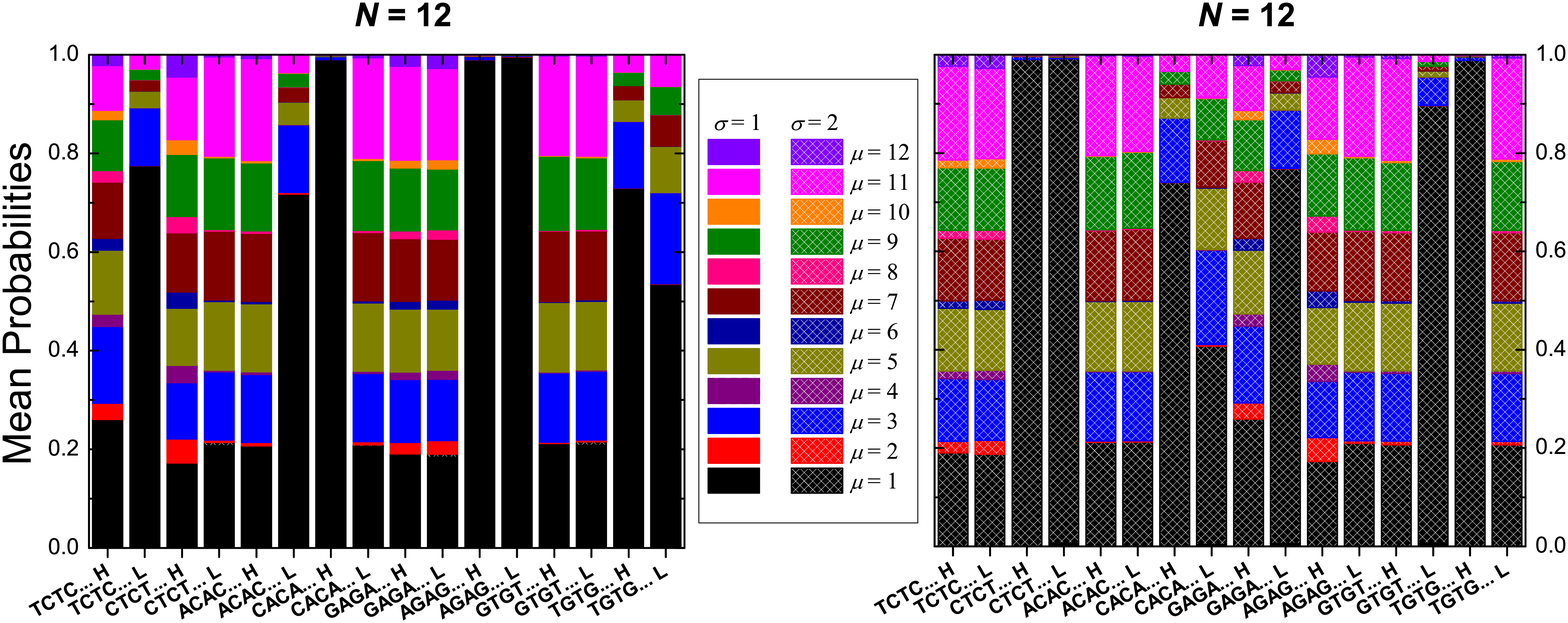}
\includegraphics[width=\textwidth]{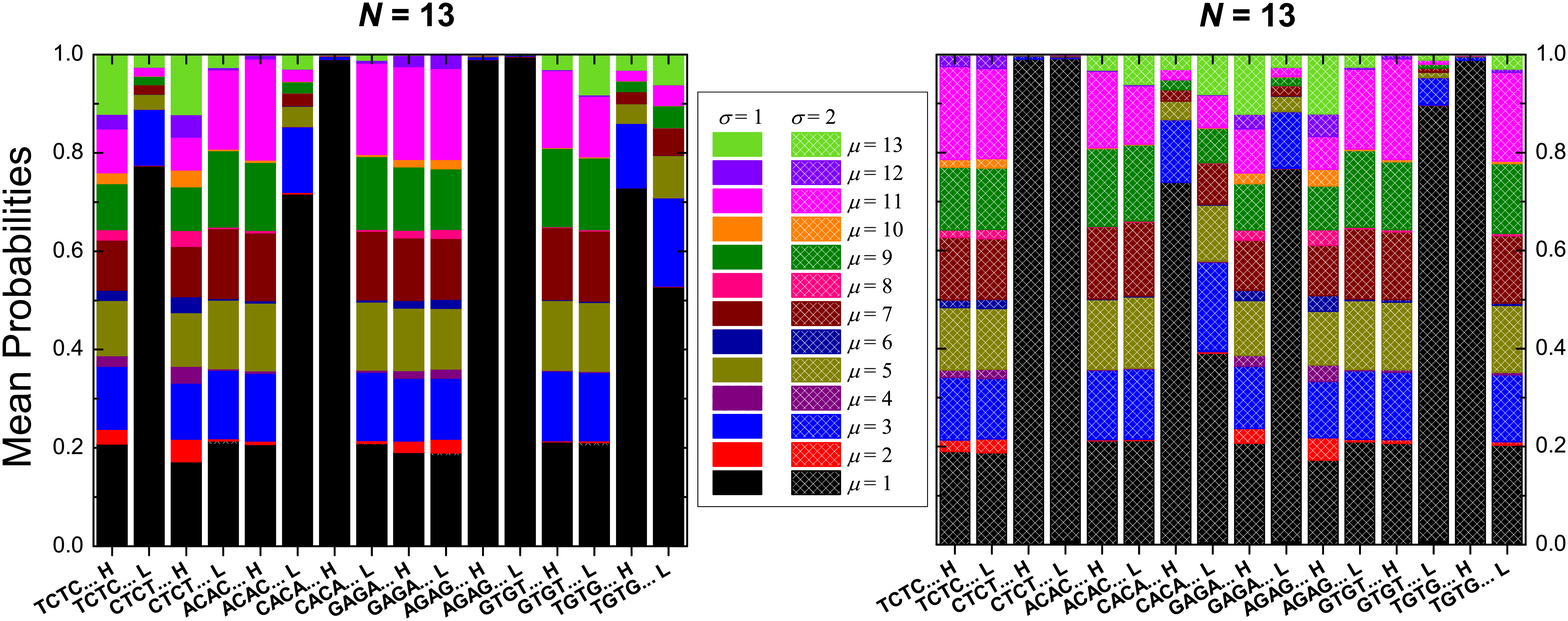}
\caption{(Color online) Type $\gamma'$ polymers.
TB II and either \textit{initial condition 1} (left) or \textit{initial condition 2} (right) , i.e.,
extra carrier initially at the first or the second base of the first base pair, respectively.
Mean over time probabilities to find an extra hole (HOMO) or electron (LUMO) at each base.
$N$ even (upper panels, here $N=12$) or $N$ odd (lower panels, here $N=13$).}
\label{fig:meanprobstypec-sb}
\end{figure}

\end{widetext}

%\pagebreak
%\clearpage

%%%%%%%%%%%%%%%%%%%%%%%%%%%%%%%%%%%%%%%%%%%%%%%%%%%%%%%%%%%%%%%%%%%%%%%%%%%%%%%%%
\subsection{\label{subsec:FrequencyContent} Charge transfer frequency content} %%
%%%%%%%%%%%%%%%%%%%%%%%%%%%%%%%%%%%%%%%%%%%%%%%%%%%%%%%%%%%%%%%%%%%%%%%%%%%%%%%%%
For TB I, for type $\alpha'$ and $\beta'$ polymers, all eigenvalues are symmetric around the on-site energy of the base pairs. Hence, the total number of frequencies involved in charge transfer is
$\frac{N^2-1}{4}$ for $N$ odd and $\frac{N^2}{4}$ for $N$ even.
For type $\gamma'$ polymers with $N$ even, the eigenvalues are symmetric around $\frac{E_1^{bp}+E_2^{bp}}{2}$,
hence the total number of frequencies is $\frac{N^2}{4}$, too.
For type $\gamma'$ polymers with $N$ odd, the eigenvalues include $E_1^{bp}$ and the total number of frequencies is $\frac{(N-1)(N+3)}{4}$.
For TB II, there are no symmetries like those mentioned for TB I, hence the total number of frequencies
for all types of polymers ($\alpha'$, $\beta'$, $\gamma'$) is $N(2N-1)$.
From Eq.~\eqref{Fourierspectra} it follows that all the palindromicity and equivalence properties presented in \S~\ref{subsec:MeanProbs} for the mean over time probabilities, $<\abs{C_j(t)}^2>$, hold for the Fourier spectra, $\abs{\mathcal{F}_j(f)}$, too. In the following Subsections, we focus on the Fourier spectra that correspond to charge transfer from the first to the last monomer i.e. on $\abs{\mathcal{F}_1(f)}$ and $\abs{\mathcal{F}_N(f)}$ for TB I, and on $\abs{\mathcal{F}_1(f)}$, $\abs{\mathcal{F}_2(f)}$, $\abs{\mathcal{F}_{2N-1}(f)}$ and $\abs{\mathcal{F}_{2N}(f)}$ for TB II.
Both TB approaches show that the frequency content is mainly in the THz domain, cf. Figs.~\ref{fig:FStypea-bp},~\ref{fig:FStypeb-bp},~\ref{fig:FStypec-bp},~\ref{fig:FStypea-sb},~\ref{fig:FStypeb-sb},~\ref{fig:FStypec-sb}.

\pagebreak
\clearpage

%%%%%%%%%%%%%%%%%%%%%%%%%%%%%%%%%%%%%%%%%%%%%%%%%%%%%%%%%%%%%%%%%%%%%
\subsubsection{\label{subsubsec:FStypea} type $\alpha'$ polymers} %%%
\begin{figure}[h!]
\centering
\includegraphics[width=7cm]{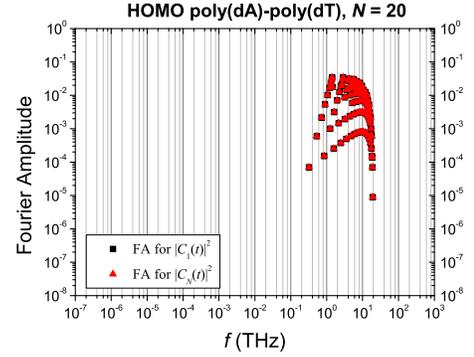}
\caption{(Color online) Type $\alpha'$ polymers, here poly(dA)-poly(dT), $N = 20$.
TB I and \textit{initial condition} (extra carrier initially at the first base pair).
Hole transfer Fourier spectra at the first and the last monomer.}
\label{fig:FStypea-bp}
\end{figure}

%%%%%%%%%%%%%%%%%%%%%%%%%%%%%%%%%%%%%%%%%%%%%%%%%%%%%%%%%%%%%%%%%%%%%
Within TB I the main frequencies are in the range $\approx$ 0.3 - 97 THz.
Within TB II, they are in the range $\approx$ 0.1 - 110 THz.
The main frequency content is between far-infrared (FIR) and mid-infrared (MIR).
As an example, we show in
Fig.~\ref{fig:FStypea-bp} (TB I and \textit{initial condition}) and
Fig.~\ref{fig:FStypea-sb} (TB II and \textit{initial condition 1} or \textit{initial condition 2}),
the Fourier spectra, at the first and the last monomer, of an extra hole in poly(dA)-poly(dT) with $N = 20$.
In Fig.~\ref{fig:FStypea-bp} we observe that the Fourier amplitudes for the first and the last monomer are approximately equal, mirroring the efficient hole transfer in poly(dA)-poly(dT), cf. also Fig.~\ref{fig:meanprobstypea-bp}.
Inspection of Fig.~\ref{fig:FStypea-sb} leads to the same conclusion.
Additionally, Fig.~\ref{fig:FStypea-sb} underlines the intra-strand character of carrier transfer and shows that initial conditions 1 and 2 lead to similar form of Fourier spectra.

%%%%%%%%%%%%%%%%%%%%%%%%%%%%%%%%%%%%%%%%%%%%%%%%%%%%%%%%%%%%%%%%%%%%
\subsubsection{\label{subsubsec:FStypeb} type $\beta'$ polymers} %%%
\begin{figure}[h!]
\centering
\includegraphics[width=7cm]{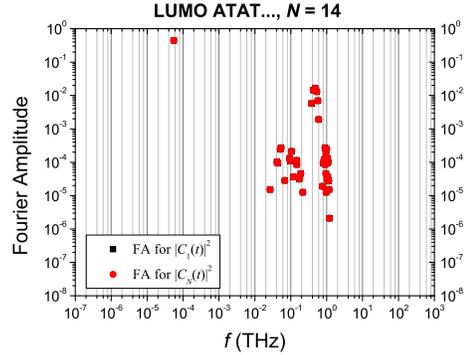}
\caption{(Color online) Type $\beta'$ polymers, here ATAT{\dots}, $N = 14$.
TB I and \textit{initial condition} (extra carrier initially at the first base pair).
Electron transfer Fourier spectra at the first and the last monomer.}
\label{fig:FStypeb-bp}
\end{figure}

%%%%%%%%%%%%%%%%%%%%%%%%%%%%%%%%%%%%%%%%%%%%%%%%%%%%%%%%%%%%%%%%%%%%
Within TB I, the main frequencies are in the range $\approx$ 0.01 - 40 THz, i.e., between microwaves (MW) and MIR.
Within TB II, they are in the range $\approx$ 0.01 - 210 THz, i.e., between the MW and near-infrared NIR.
As an example, we show in
Fig.~\ref{fig:FStypeb-bp} (TB I and \textit{initial condition}) and
Fig.~\ref{fig:FStypeb-sb} (TB II and \textit{initial condition 1} or \textit{initial condition 2}),
the Fourier spectra, at the first and the last monomer, of an extra electron in ATAT{\dots} with $N = 14$.
In Fig.~\ref{fig:FStypeb-bp} we observe that the Fourier amplitudes for the first and the last monomer are approximately equal, mirroring the finally large electron transfer in ATAT{\dots} for $N$ even, cf. also Fig.~\ref{fig:meanprobstypeb-bp}.
However, this large transfer is very slow, its main frequency is very small but with a large amplitude.
The same conclusions can be drawn from Fig.~\ref{fig:FStypeb-sb}, where we can additionally observe the inter-strand character of charge transfer and that initial conditions 1 and 2 lead to similar form of Fourier spectra.

%%%%%%%%%%%%%%%%%%%%%%%%%%%%%%%%%%%%%%%%%%%%%%%%%%%%%%%%%%%%%%%%%%%%%
\subsubsection{\label{subsubsec:FStypec} type $\gamma'$ polymers} %%%
\begin{figure} [h!]
\centering
\includegraphics[width=7cm]{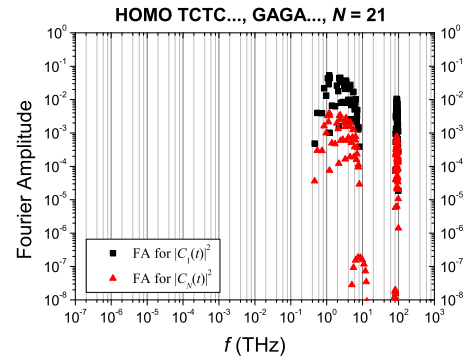}
\caption{(Color online) Type $\gamma'$ polymers, here TCTC{\dots}, $N = 21$.
TB I and \textit{initial condition} (extra carrier initially at the first base pair).
Hole transfer Fourier spectra at the first and the last monomer.}
\label{fig:FStypec-bp}
\end{figure}

%%%%%%%%%%%%%%%%%%%%%%%%%%%%%%%%%%%%%%%%%%%%%%%%%%%%%%%%%%%%%%%%%%%%%
Within TB I, the main frequencies are in the range $\approx 0.4$ GHz - $ 40$ THz, i.e., between radiowaves (RW) and MIR.
Within TB II, they are in the range $\approx$ 0.02 - 190 THz, i.e., between MW and FIR.
As an example, we show in
Fig.~\ref{fig:FStypec-bp} (TB I and \textit{initial condition}) and
Fig.~\ref{fig:FStypec-sb} (TB II and \textit{initial condition 1} or \textit{initial condition 2}),
the Fourier spectra, at the first and the last monomer, of an extra hole in TCTC{\dots} with $N = 21$.
In Fig.~\ref{fig:FStypec-bp} we observe that the Fourier amplitudes for the first monomer are much larger than the ones for the last monomer, mirroring the inefficient hole transfer in TCTC{\dots} for $N$ odd, cf. also Fig.~\ref{fig:meanprobstypec-bp}.
In Fig.~\ref{fig:FStypec-sb}, we can additionally observe the intra-strand character of charge transfer and that initial conditions 1 and 2 lead to somehow different form of Fourier spectra, initial condition 1 being more efficient than initial condition 2 for hole transfer, cf. also Figs.~\ref{fig:meanprobstypec-bp}~and~\ref{fig:meanprobstypec-sb}.

\pagebreak

\begin{widetext}

\begin{figure}[h!]
\centering
\includegraphics[width=14cm]{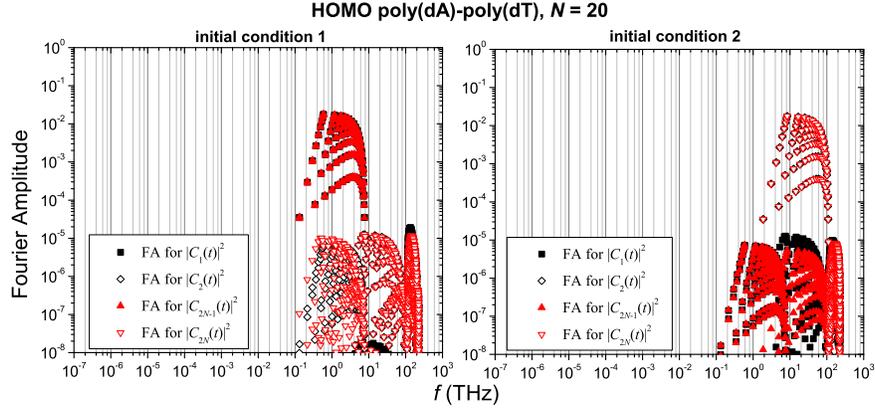}
\caption{(Color online) Type $\alpha'$ polymers, here poly(dA)-poly(dT), $N = 20$.
TB II and either \textit{initial condition 1} (left) or \textit{initial condition 2} (right).
Hole transfer Fourier spectra at the first and the last monomer.}
\label{fig:FStypea-sb}
\end{figure}

\begin{figure}[h!]
\centering
\includegraphics[width=14cm]{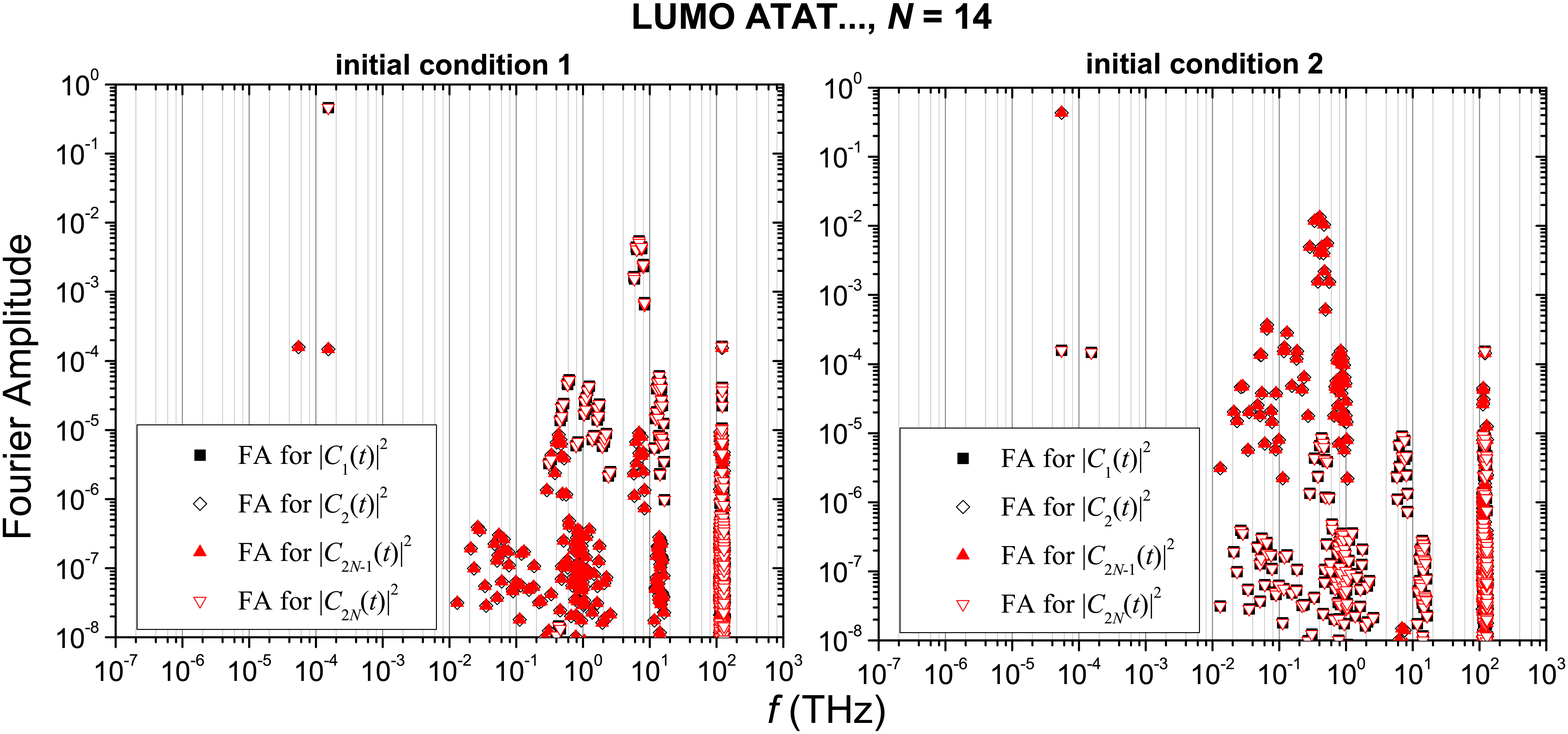}
\caption{(Color online) Type $\beta'$ polymers, here ATAT{\dots}, $N = 14$.
TB II and either \textit{initial condition 1} (left) or \textit{initial condition 2} (right).
Electron transfer Fourier spectra at the first and the last monomer.}
\label{fig:FStypeb-sb}
\end{figure}

\begin{figure} [h!]
\centering
\includegraphics[width=14cm]{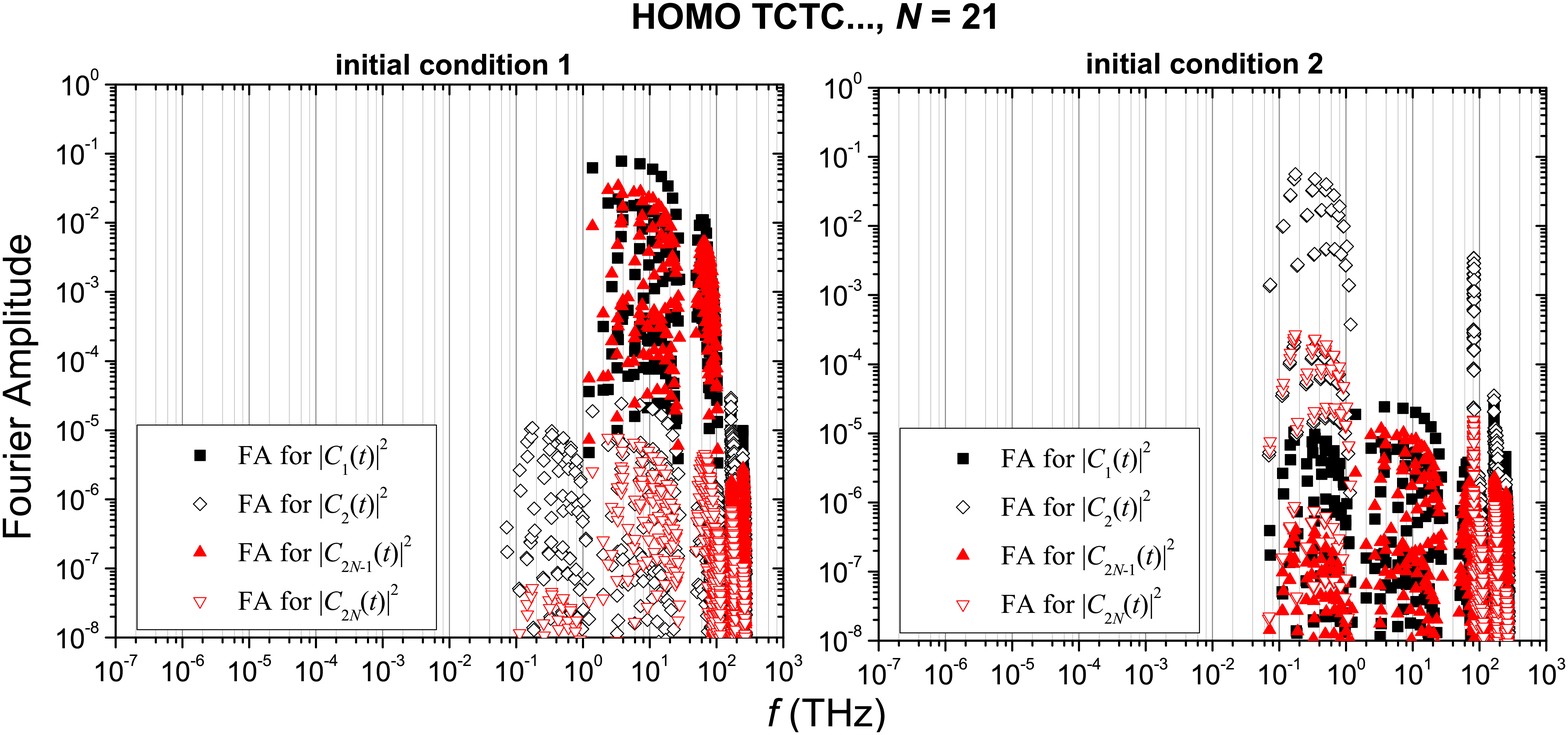}
\caption{(Color online) Type $\gamma'$ polymers, here TCTC{\dots}, $N = 21$.
TB II and either \textit{initial condition 1} (left) or \textit{initial condition 2} (right).
Hole transfer Fourier spectra at the first and the last monomer.}
\label{fig:FStypec-sb}
\end{figure}

\end{widetext}

%\pagebreak
%\clearpage

%%%%%%%%%%%%%%%%%%%%%%%%%%%%%%%%%%%%%%%%%%%%%%%%%%%%%%%%%%%%%%%%%%%%%%%%%
\subsection{\label{subsec:meantransferrates} Pure mean transfer rates} %%
%%%%%%%%%%%%%%%%%%%%%%%%%%%%%%%%%%%%%%%%%%%%%%%%%%%%%%%%%%%%%%%%%%%%%%%%%
In the following subsections we focus on pure mean transfer rates between the first and the last monomer, either within TB I or within TB II.

%%%%%%%%%%%%%%%%%%%%%%%%%%%%%%%%%%%%%%%%%%%%%%%%%%%%%%%%%%%%%%%%%%%%
\subsubsection{\label{subsubsec:ktypea} type $\alpha'$ polymers} %%%
%%%%%%%%%%%%%%%%%%%%%%%%%%%%%%%%%%%%%%%%%%%%%%%%%%%%%%%%%%%%%%%%%%%%
As a characteristic example, we present in Fig.~\ref{fig:ktypea} the hole pure mean transfer rates for poly(dG)-poly(dC),
from the first to the last monomer, either within TB I or within TB II. Specifically,
(I)  for TB I  we illustrate $k_{1,N}$ on the left panel, and
(II) for TB II we illustrate $k_{1,2} = k_{2,1}$, $k_{1,2N-1} = k_{2,2N}$ and $k_{1,2N} = k_{2,2N-1}$ on the right panel.
We have already noticed in \S~\ref{subsubsec:meanProbstypea} that, within TB II, carrier transfer is almost exclusively of  intra-strand character. Hence, within TB II, $k_{1,2N-1} = k_{2,2N}$ are the largest transfer rates.
Comparing $k_{1,N}$ for TB I with $k_{1,2N-1} = k_{2,2N}$ for TB II, we observe an excellent agreement, both qualitatively and quantitatively. Within TB II, the intra-base-pair rates $k_{1,2} = k_{2,1}$ are small and the inter-strand rates $k_{1,2N} = k_{2,2N-1}$ insignificant. Increasing $N$, the intra-strand transfer rates $k_{1,2N-1} = k_{2,2N}$ decrease reaching gradually the level of the the intra-base-pair rates $k_{1,2} = k_{2,1}$, at which point, finally, charge transfer along the polymer is insignificant. Increasing $N$, the insignificant inter-strand rates $k_{1,2N} = k_{2,2N-1}$ also gradually decrease further.

%%%%%%%%%%%%%%%%%%%%%%%%%%%%%%%%%%%%%%%%%%%%%%%%%%%%%%%%%%%%%%%%%%%
\subsubsection{\label{subsubsec:ktypeb} type $\beta'$ polymers} %%%
%%%%%%%%%%%%%%%%%%%%%%%%%%%%%%%%%%%%%%%%%%%%%%%%%%%%%%%%%%%%%%%%%%%
We have already mentioned (cf. \S~\ref{subsubsec:meanProbstypeb}~and~\ref{subsubsec:FStypeb}) that both TB approaches predict that for some cases of type $\beta'$ polymers, for $N$ even, the carrier is transferred at a large percentage to the last monomer but the transfer is very slow. Such a case is presented in Fig.~\ref{fig:ktypeb}.
Specifically, we show the electron pure mean transfer rates for ATAT{\dots},
from the first to the last monomer, either within TB I or within TB II. Specifically,
(I)  for TB I,  we illustrate $k_{1,N}$ (left), and
(II) for TB II, we illustrate the largest transfer rates (right).
We have already demonstrated in \S~\ref{subsubsec:meanProbstypeb} that, within TB II, the extra carrier is transferred almost exclusively crosswise, through identical bases. Hence, for TB II, the largest transfer rates are
$k_{1,2N-1}$ and $k_{2,2N}$ for $N$ odd, and
$k_{1,2N}$ and $k_{2,2N-1}$ for $N$ even.
We depict these largest transfer rates in Fig.~\ref{fig:ktypeb} (right).
In other cases of type $\beta'$ polymers the pure mean transfer rates fall over $N$ in a different manner, somehow similar to the behavior of type $\gamma'$ polymers, which is shown in \S~\ref{subsubsec:ktypec}.

%%%%%%%%%%%%%%%%%%%%%%%%%%%%%%%%%%%%%%%%%%%%%%%%%%%%%%%%%%%%%%%%%%%%
\subsubsection{\label{subsubsec:ktypec} type $\gamma'$ polymers} %%%
%%%%%%%%%%%%%%%%%%%%%%%%%%%%%%%%%%%%%%%%%%%%%%%%%%%%%%%%%%%%%%%%%%%%
As a characteristic example, we present in Fig.~\ref{fig:ktypec} the hole pure mean transfer rates for TCTC{\dots},
from the first to the last monomer, either within TB I or within TB II. Specifically,
(I)  for TB I  we illustrate $k_{1,N}$ on the left panel, and
(II) for TB II we illustrate $k_{1,2N-1}$ and $k_{2,2N}$ on the right panel.
We have already mentioned in \S~\ref{subsubsec:meanProbstypec} that, within TB II, the extra carrier is transferred almost exclusively through the strand it was initially placed at, i.e., for type $\gamma'$ polymers the charge transfer is mainly of intra-strand character.
Hence, for TB II, we show $k_{1,2N-1}$ and $k_{2,2N}$ which are the largest transfer rates.
We have demonstrated \S\ref{subsubsec:ktypea} that, for type $\alpha'$ polymers, $k_{1,2N-1} = k_{2,2N}$.
As shown in Fig.~\ref{fig:ktypec}, this does not hold for type $\gamma'$ polymers.

%%%%%%%%%%%%%%%%%%%%%%%%%%%%%%%%%%%%%%%%%%%%%%%%%%%%%%%%%%%%%%%%%%%%%%%%
\subsubsection{\label{subsubsec:kfits} Pure mean transfer rate fits} %%%
%%%%%%%%%%%%%%%%%%%%%%%%%%%%%%%%%%%%%%%%%%%%%%%%%%%%%%%%%%%%%%%%%%%%%%%%
Finally, we compare the results of our two TB approaches by performing the exponential fits $k = k_0 e^{-\beta d}$ and $k = A + k_0 e^{-\beta d}$, where $d = (N-1)\times 3.4$ {\AA} is the charge transfer distance, as well as the power-law fit, $k = k_{0}' N^{-\eta}$. Our results for TB I have already been presented in Figs.~8-9 of Ref.~\cite{LChMKTS:2015}. For TB II, we again focus on the pure mean transfer rates between the bases of the initial and the final monomer for which carrier transfer is significant.
The conclusions are similar to those within TB I. The fits are considerably improved if polymers with $N$ odd and $N$ even are fitted separately. Moreover, the power-law fits are generally better, in terms of correlation coefficients. Our results for the exponent $\eta$ of the power-law fits, within TB II, are presented in Fig.~\ref{fig:PLfitssb}. Our results confirm the statement that the fall of $k$ as a function of $N$ becomes generally steeper as the intricacy of the energy structure increases, i.e., from type $\alpha'$ to type $\beta'$ and further to type $\gamma'$ polymers~\cite{LChMKTS:2015}. This conclusion also holds for the exponential fits, which are not presented here. Furthermore, both TB I and TB II show that there is perfect agreement between our results for $\beta$ and $\eta$ for all type $\alpha'$ polymers. This leads to the conclusion that although the interaction strength (as reflected in the hopping integrals) is different in each case of type $\alpha'$ polymers leading to different values of $k$, the way $k$ falls over $N$ or $d$ is the same.

\begin{widetext}

\begin{figure} [h]
\centering
\includegraphics[width=7cm]{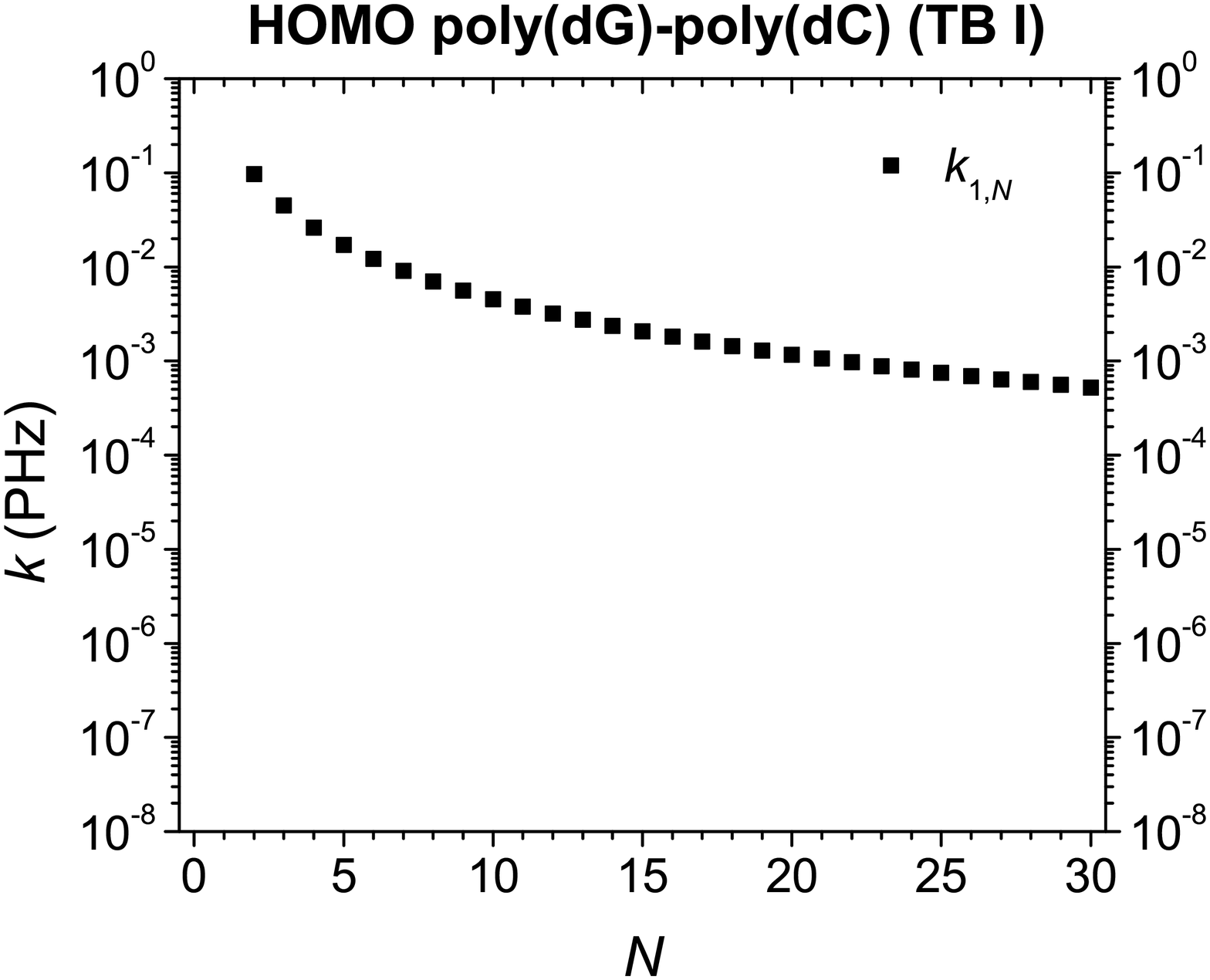}
\includegraphics[width=7cm]{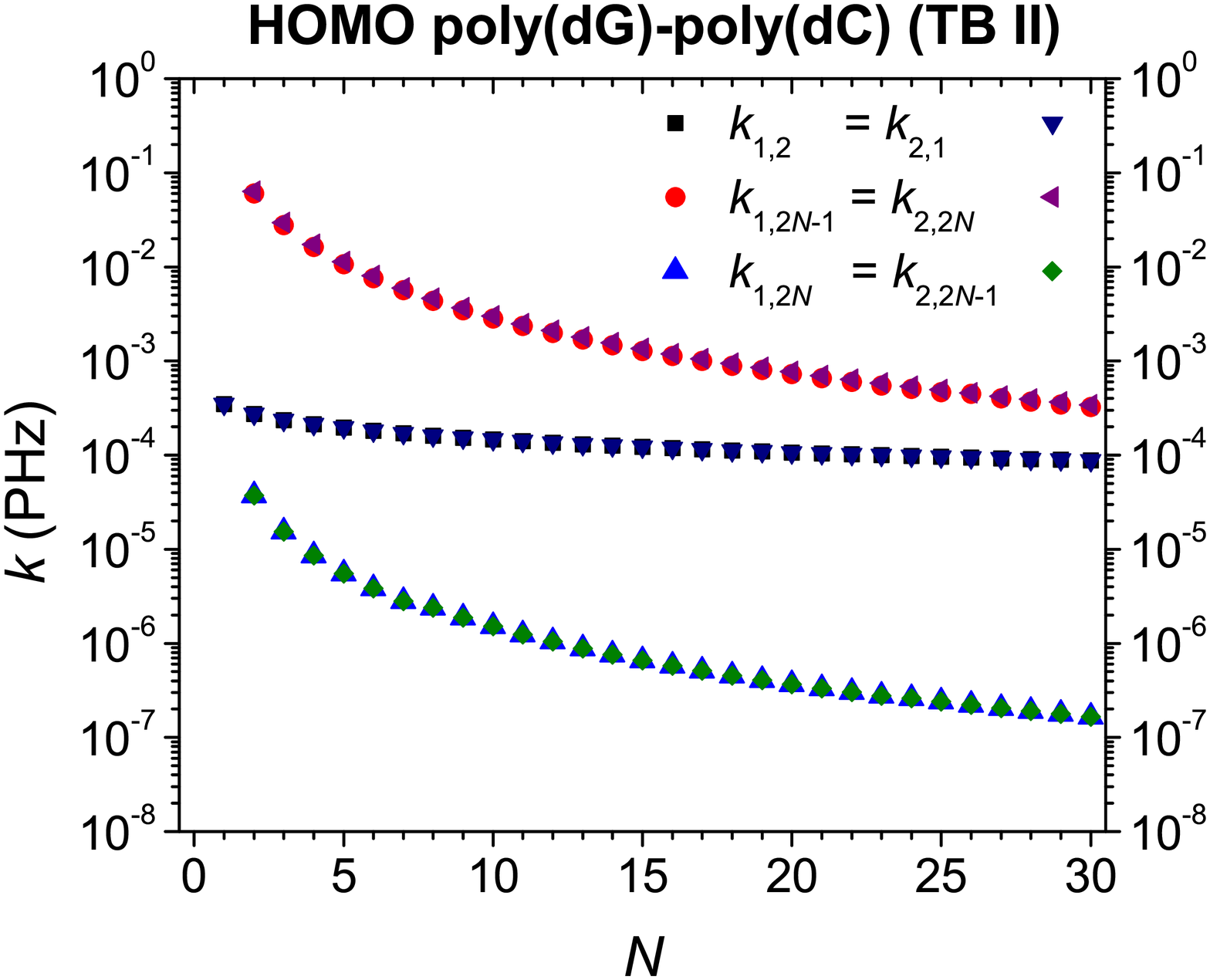}
\caption{(Color online) Type $\alpha'$ polymers, here poly(dG)-poly(dC).
Hole pure mean transfer rates
(I) $k_{1,N}$ for TB I (left), and
(II) $k_{1,2} = k_{2,1}$, $k_{1,2N-1} = k_{2,2N}$ and $k_{1,2N} = k_{2,2N-1}$ for TB II (right).}
\label{fig:ktypea}
\end{figure}

\begin{figure}[h!]
\centering
\includegraphics[width=7cm]{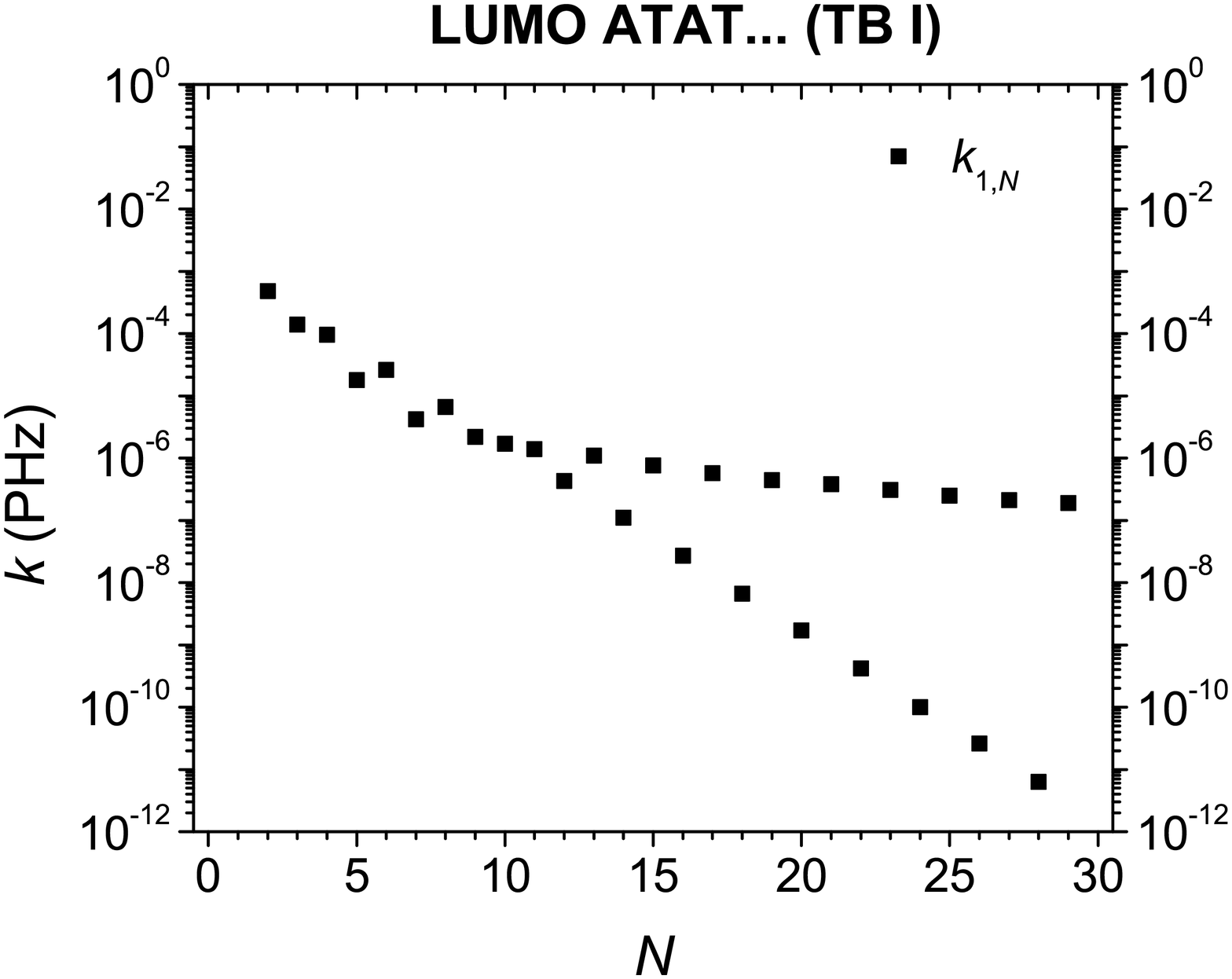}
\includegraphics[width=7cm]{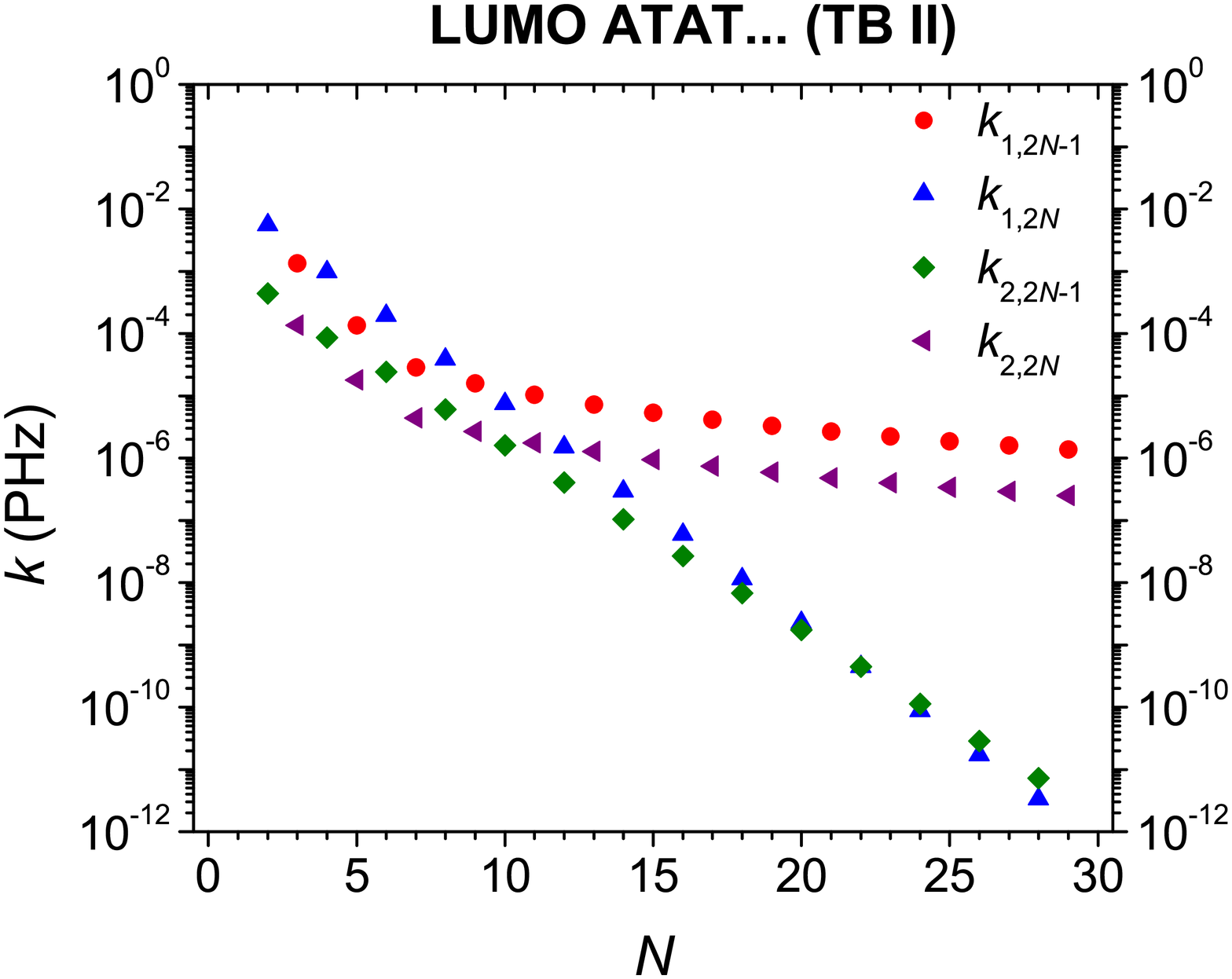}
\caption{(Color online) Type $\beta'$ polymers, here is a case where, for $N$ even, the carrier is transferred at a large percentage to the last monomer but the transfer is very slow: Electron pure mean transfer rates in ATAT{\dots},
either $k_{1,N}$ within TB I (left) or the largest transfer rates in TB II, i.e., $k_{1,2N-1}$ for $N$ odd, $k_{1,2N}$ for $N$ even, $k_{2,2N}$ for $N$ odd, $k_{2,2N-1}$ for $N$ even.}
\label{fig:ktypeb}
\end{figure}

\begin{figure}[h]
\centering
\includegraphics[width=7cm]{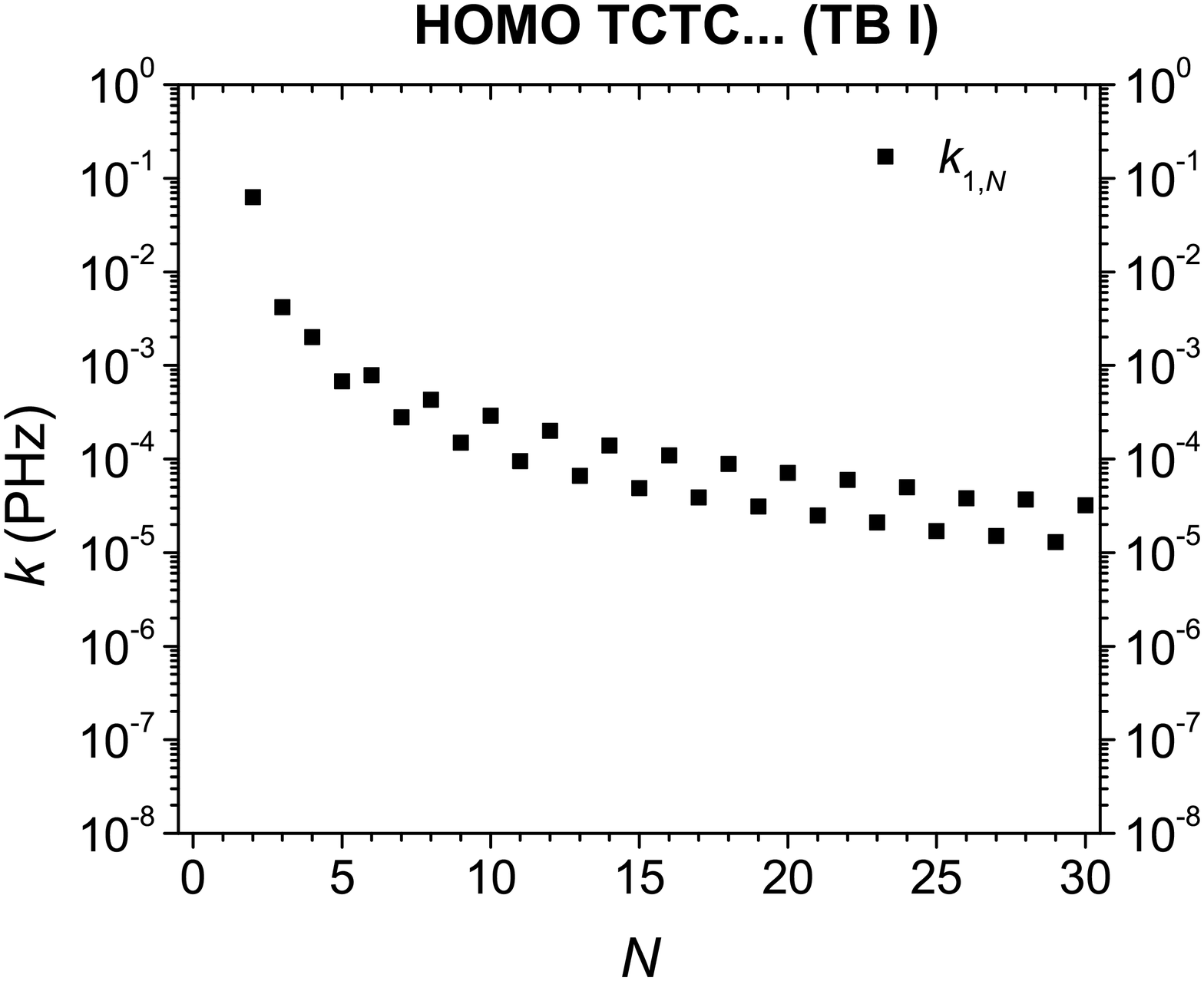}
\includegraphics[width=7cm]{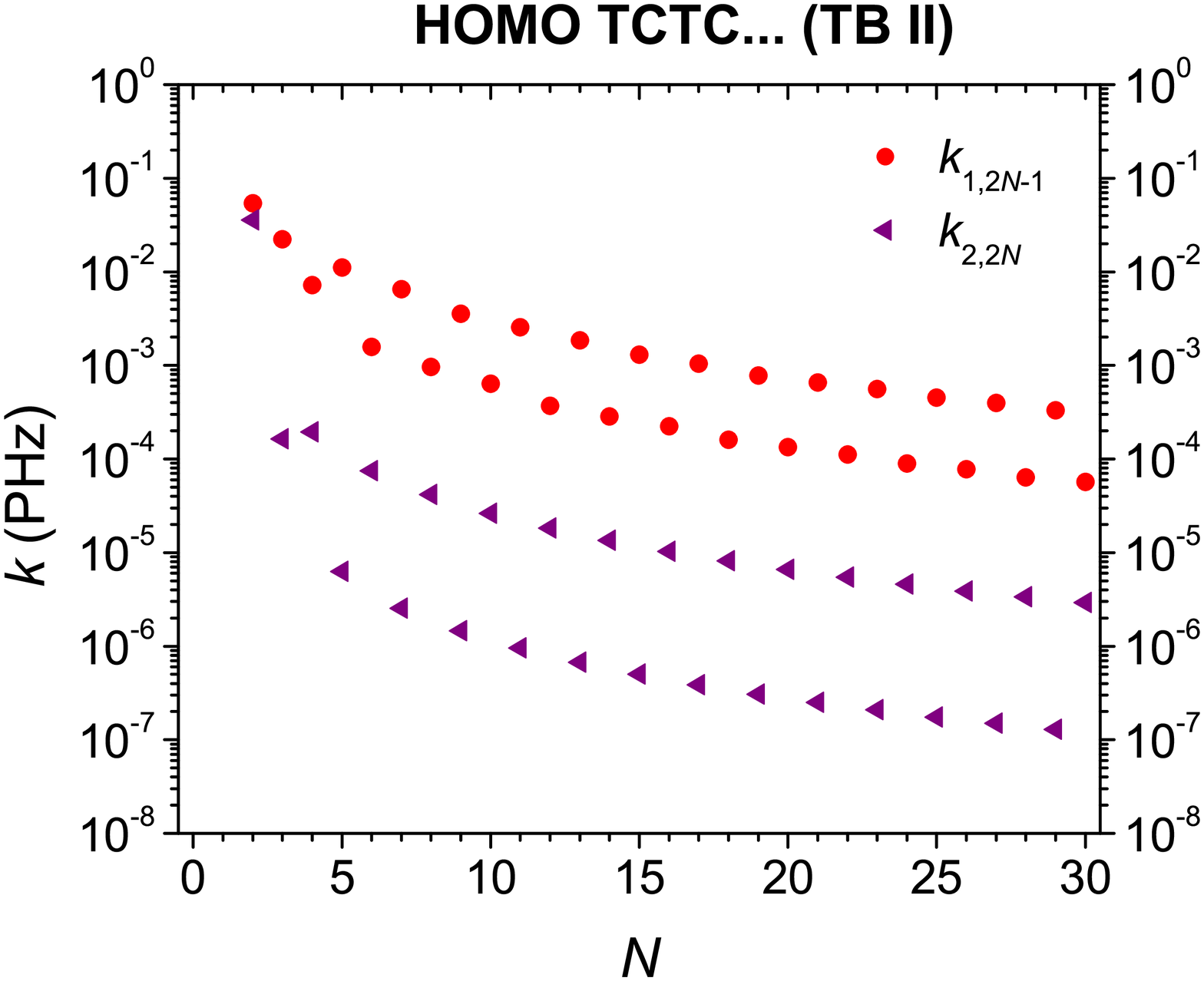}
\caption{(Color online) Type $\gamma'$ polymers, here TCTC{\dots}.
Hole pure mean transfer rates $k_{1,N}$ for TB I (left), and $k_{1,2N-1}$, $k_{2,2N}$ for TB II (right).}
\label{fig:ktypec}
\end{figure}

\end{widetext}

\clearpage

%%%%%%%%%%%%%%%%%%%%%%%%%%%%%%%%%%%%%%%%%%%%%%
\section{\label{sec:Conclusion} Conclusion} %%
%%%%%%%%%%%%%%%%%%%%%%%%%%%%%%%%%%%%%%%%%%%%%%
We employed two Tight-Binding approaches to examine time-independent and time-dependent aspects of the electronic structure and carrier transfer in B-DNA monomer polymers (type $\alpha'$) and dimer polymers (type $\beta'$ and type $\gamma'$).
We used a simplistic wire model (TB I) where a carrier is initially located at a base pair (called also a monomer in this article) and then moves to the next or to the previous base pair, as well as a more detailed extended ladder model (TB II) where the carrier is initially located at a base and then moves to all possible neighboring bases including diagonally located ones.
The inclusion of diagonal hoppings is crucial for type $\beta'$ polymers where carrier transfer is mainly of inter-strand character.
The time-dependent and the time-independent problems involve diagonalization of matrices with
matrix dimension $MD$ = $N$ for TB I and $MD$ = $2N$ for TB II.
The two TB approaches give coherent, complementary aspects of electronic properties and charge transfer in B-DNA monomer polymers and dimer polymers.

For the time-independent problem, we studied the HOMO and the LUMO eigenspectra and the occupation probabilities, the Density of States and the HOMO-LUMO gap.
The upper (lower) subband of the HOMO (LUMO) eigenspectrum calculated with TB II corresponds to the band calculated with TB I.
The occupation probabilities within TB I and TB II show various degrees of palindromicity and eigenspectrum (in)dependence of the probabilities to find the carrier at a site.
The DOS displays nice van Hove singularities at the (sub)band edges, while the numerically calculated DOS for simple cases agrees with the analytical solution.
As expected, the polymer HOMO-LUMO gaps are smaller than the HOMO-LUMO gaps of the two possible monomers, reaching a level of 3.4 to 3.0 eV. The smallest HOMO-LUMO gaps occur for type $\gamma'$ polymers.

For the time-dependent problem we investigated the mean over time probabilities to find the carrier at each site (base pair for TB I and base for TB II), the Fourier spectra and the pure mean transfer rates from a certain site to another.
The mean over time probabilities illustrate clearly the basically intra-strand character of carrier transfer in type $\alpha'$ and type $\gamma'$ polymers. However, while in type $\alpha'$ polymers the carrier moves successively through all bases of the same strand, in type $\gamma'$ polymers the carrier moves through the bases that are identical with the one it was initially placed at, i.e., it moves through the same strand from the one or the other base of the first monomer to the identical base of the third monomer, and so forth. Carrier transfer is basically of inter-strand character in type $\beta'$ polymers.
The Fourier spectra give us a nice representation of the frequency content of charge transfer.
Both TB approaches show that this frequency content is mainly in the THz domain, the details depend on the type of polymers and the TB approach used.
The pure mean transfer rates $k$ show both how fast carrier transfer is and how much of the carrier is transferred from the initial site to the final site. The $k(N)$ fits are considerably improved if polymers with $N$ odd and $N$ even are fitted separately. Additionally, the power-law fits are generally better, in terms of correlation coefficients.
Our results confirm the statement that the fall of $k$ as a function of $N$ becomes generally steeper as the intricacy of the energy structure increases, i.e., from type $\alpha'$ to type $\beta'$ and further to type $\gamma'$ polymers.

\begin{widetext}

\begin{figure} [h]
\centering
\includegraphics[width=\textwidth]{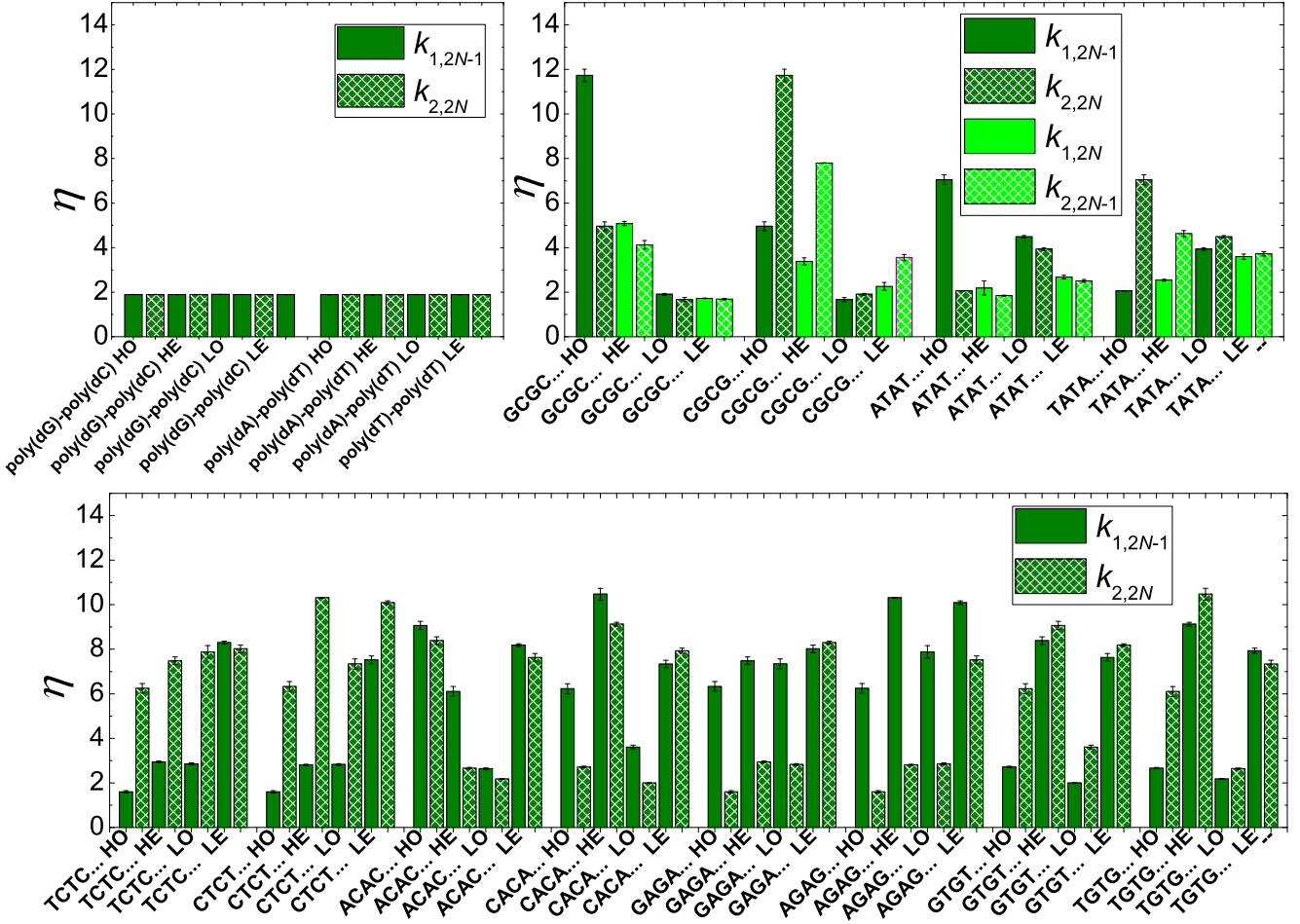}
\caption{(Color online) The exponent $\eta$ of the power-law fits, $k = k_{0}' N^{-\eta}$, within TB II.}
\label{fig:PLfitssb}
\end{figure}

\end{widetext}

%\clearpage

\end{document}